\documentclass[aps,pre,showpacs,groupedaddress,showkeys,amsmath,amssymb]{revtex4}
\usepackage{amsmath}
\usepackage{amssymb}
\usepackage{color}
\usepackage[T1]{fontenc}
\usepackage{graphicx}
\usepackage{subfigure}
\usepackage{theorem}
\usepackage{enumerate}

\newcommand{\Om}{\Omega}

\newcommand{\I}{\mathbb{I}}

\newcommand{\eps}{\varepsilon}
\newcommand{\vp}{\varphi}
\renewcommand{\div}{\operatorname{div}}

\newcommand{\rot}{\nabla\!\times\!}

\def\beq{\begin{equation}}
\def\eeq{  \end{equation}}
\def\gp{\bnabla\phi}
\def\gvp{\bnabla\vp}

\def\gd{\nabla\delta}
\def\ngp{|\bnabla\phi|}
\def\ngvp{|\bnabla\vp|}

\def\ze{\frac1\eps\zeta(\frac\vp\eps)}
\def\Om{\Omega}
\newcommand{\zep}{\frac1{\eps^2}\zeta'(\frac{\vp}{\eps})}
\newcommand{\prpo}{\mathbb{P}_{\gvp^\perp}}
\newcommand{\rr}{\rho_{\textrm{ref}}}
\newcommand{\mr}{\eta_{\textrm{ref}}}
\def\bnabla{\nabla}
\def\bx{x}
\def\bu{u}
\def\bF{F}
\def\bv{v}
\newcommand{\kp}{c(\vp)}
\newcommand{\bfs}[1]{\boldsymbol{#1}}

\begin{document}
\title{Comparison between advected-field and level-set methods in the study of vesicle dynamics}
\date{}
\author{E. Maitre$^1$, C. Misbah$^2$, P. Peyla$^2$ and A. Raoult$^3$}
\affiliation{$^1$ Laboratoire Jean Kuntzmann, Universit\'e Joseph
Fourier and CNRS,  B. P. 53 38041 Grenoble Cedex 9,
France\\$^2$Laboratoire de Spectrom\'etrie Physique, UMR, 140 avenue
de la physique, Universit\'e Joseph Fourier, and CNRS, 38402 Saint
Martin d'Heres, France\\ $^3$ Laboratoire MAP5, (UMR CNRS 8145),
Université Paris Descartes, Paris, France}

\begin{abstract}
Phospholipidic membranes and vesicles constitute a basic element in
real biological functions. Vesicles are viewed as a model system to
mimic  basic viscoelastic behaviors of some cells, like red blood
cells. Phase field and level-set models are powerful tools to tackle
dynamics of membranes and their coupling to the flow. These two methods are somewhat similar, but to date
no bridge between them has been made. This is a first focus of this paper.
Furthermore,  a constitutive viscoelastic law is derived for the
composite fluid: the ambient fluid and the membranes. We  present
two different approaches to deal with the membrane local
incompressibility, and point out differences. Some
numerical results following from the level-set approach  are presented.
\end{abstract}
 \maketitle
\section{Introduction}
\begin{figure}
\begin{center}
\includegraphics[width=.4\linewidth]{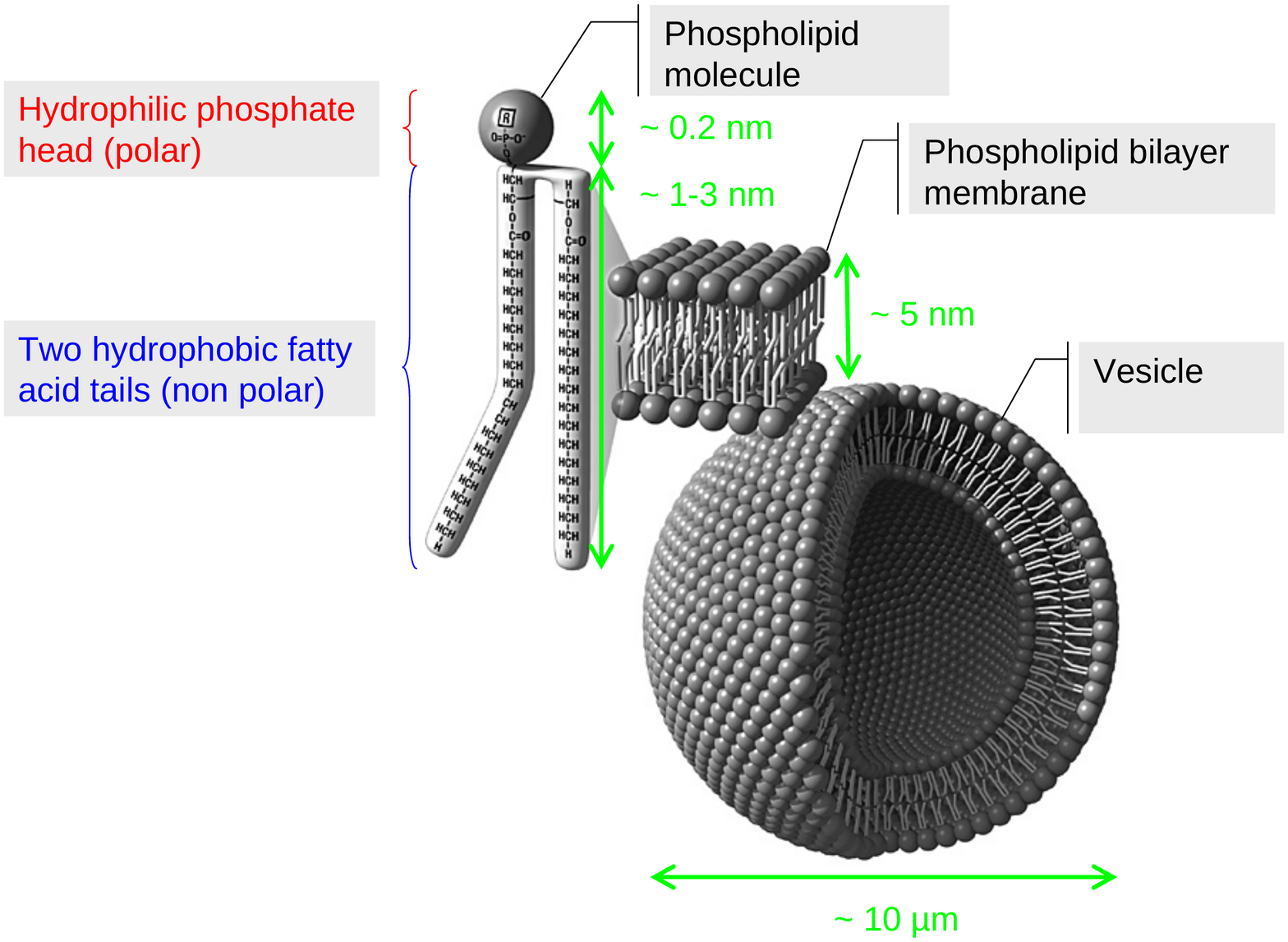}
  \caption{A cartoon showing a vesicle and the molecular structure of its membrane. Picture taken from the web site of the Nasa Astrobiology Institute: \textit{http://astrobiology.nasa.gov/nai/}}. \label{fig_membrane}
\end{center}
\end{figure}

Vesicles are closed membranes which are suspended in an aqueous
solution. The membrane is made of a bilayer of phospholipid
molecules. These molecules have a polar hydrophilic head which
points towards the solvent, and hydrophobic tails which point
towards the interior of the membrane (Fig.\ref{fig_membrane}). Both
at room and physiological temperatures the bilayer is a two dimensional
incompressible fluid.

\begin{figure}
\begin{center}
\includegraphics[width=.7\linewidth]{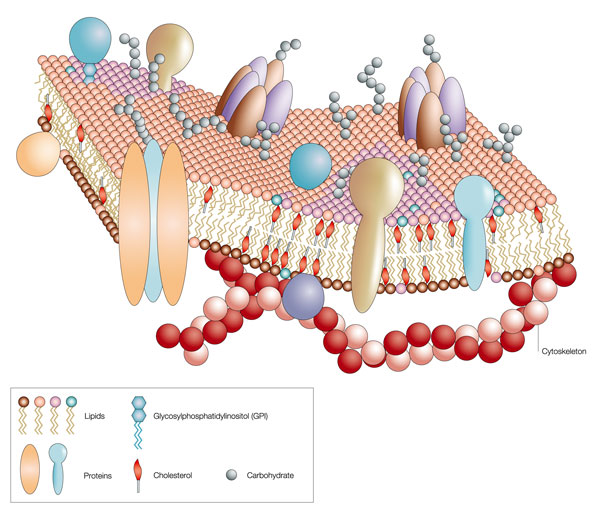}
  \caption{A schematic view of a red blood cell membrane. Besides the bilayer of
   phospholipids, there is a network
  of proteins (called also cytoskeleton) and other proteins. }. \label{rbc}
\end{center}
\end{figure}

Human red blood cells (RBCs) are among the simplest animal cells. They are made, like
vesicles, of phospholipid bilayer, plus a protein network
(called spectrin), known also under the generic name of cytoskeleton (Fig.\ref{rbc}). The internal
solution of RBC is made of a hemoglobin solution (a newtonian
fluid). The RBC is devoid of a nucleus and organella, and this is why it may be viewed as a simple cell. Its main function consists in oxygen supply to tissues. It is  hoped that vesicles may represent a simplistic
starting point to understand viscoelastic properties, dynamics and
rheology of bio-fluids, such as blood.

The problem of vesicles at equilibrium (equilibrium shapes) is now
fairly understood \cite{Seifert1997}. The
study of vesicles under non-equilibrium conditions  has constituted a major focus in recent years \cite{Kraus1996,Haas1997,Seifert1999,Cantat1999,Abkarian2002,Biben2003,Beaucourt2004,Rioual2004,Noguchi2004,Kantsler2005,Noguchi2005a,Kantsler2006,Mader2006,Misbah2006,Vlahovska2007,Noguchi2007,Lebedev2008,Danker2007,Danker2007a,Kessler2008,Finken2008,Vitkova2008,Kantsler2008,Coupier2008,McWhirter2009,Deschamps2009,Danker2009a} (see recent review \cite{Vlahovska2009}).
 Vesicles under flow have revealed
 several fascinating non-equilibrium behaviors
(tank-treading, tumbling, vacillating-breathing--aka swinging, trembling, and so on). RBCs exhibit also
similar kinds of dynamics (see review \cite{Vlahovska2009}). This  research activity knows
nowadays an increasing interest in the field of physics, mechanics, applied
mathematics, engineering science, and so on. This field   has known during the past decade   analytical,
numerical and experimental progresses (see review \cite{Vlahovska2009}). Even at zero Reynolds number,
where the hydrodynamics equations are linear (the Navier-Stokes
equations reduce to the Stokes equations), the problem remains
highly nontrivial due to the fluid/membrane coupling. This coupling
triggers nonlinearity (inherent to any moving boundary problem, even
if the bulk equations are linear), and non-locality (a disturbance of
the membrane shape at some point is felt by other regions of the
membranes due to the fact that hydrodynamics have long range effects; very
much like Coulomb interactions in electrostatics). In mathematical
terms, the velocity field in the fluid can be integrated out in
favor of a closed nonlinear integro-differential equations. This is
the so-called boundary integral formulation based on the use of the Greens
function technique \cite{Pozrikidis1992}. This method has been used for vesicles by several groups \cite{Kraus1996,Cantat1999,Veerapaneni2009}. This method, despite its efficiency and precision, has some limitations. For example, in a
situation where the internal (or external) fluid is non newtonian,
and/or if the non zero Reynolds number limit (a situation
encountered for blood flows in veins and arteries, for example) has to be treated, then the resulting nonlinearities
of the fluid equations clearly rule out the use of the Greens function techniques.  We have thus to resort to other methods.

Phase-field and level-set approaches can, in principle, handle the
above problems  quite naturally. Phase field methods have known some
popularity in the scientific  community, and have been used  in various  topics where  interfacial phenomena are present (crystal
growth, fluid/fluid interfaces, and so on). More recently, a
phase-field method has been adapted to vesicles \cite{Biben2003,Biben2005}. Later, other phase-field formulations have been presented for the same system  \cite{Du2005,Campelo2006,Jamet2008a}.

In classical
interfacial problems (crystal growth, vapor/liquid phase transitions,
fracture...), the total length of the interface can grow (or shrink) without bound. In contrast, for membranes a new important ingredient comes into play:
the membrane is a two dimensional incompressible fluid, and thus its area does not change in the course of time. More precisely,  not only should  the total area  remain constant in the course of time, but also the area must be conserved locally. This
question is not easy to deal with in practice (a reason why only global area conservation have been imposed in some phase-field models \cite{Du2005,Campelo2006}). Like the phase-field, the level-set approach treats the problem in the same spirit, but the precise formulations differ in each case, as we shall see here. How does the level-set formulation  compare with the phase-field one is a question which has not yet  been treated in the literature, and especially regarding the problem of vesicles and membranes. This paper addresses this question as a first focus.

For the sake of completeness, it is worthwhile to cite other  alternative methods which have been adapted for the  study of biological membranes. These are (i) techniques based on dynamically triangulated models
\cite{Gompper1997} or particle-based mesoscale solvent, multi-particle
collision dynamics \cite{Malvanets1999}, or their combination
\cite{Noguchi2004}, (iv) the so-called immersed boundary method
\cite{Peskin1977}, used to model some features of red blood cells \cite{Eggleton1998}, and
 (v) Lattice-Boltzmann methods \cite{Dupin2007}.

In this paper we shall draw a parallel between the phase-field and
level-set methods. It will be shown that, despite their apparent
differences, the two methods have a quite number of links.  The level-set that we shall present here will be compared to the phase-field models previously  introduced in two and three dimensions \cite{Biben2003,Biben2005}. We shall make a bridge between the two methods. We shall also give here a derivation of the postulated equation in Refs. \cite{Biben2003,Biben2005} regarding the treatment of the local membrane incompressibility.
Finally, we shall show that the membrane force derived in
\cite{Biben2003,Biben2005} can be written as a divergence of some rank two tensor, and
this will allow us  to write a closed form for  the constitutive law of
the fluid/membrane system. This law exhibits a viscoelastic
behavior. We shall then briefly discuss the numerical method used to solve the level-set equations. Some numerical results will be presented.

The scheme of the paper is as follows. In section \ref{preliminaries} we present some essential preliminaries. In section \ref{phase-field}   we recall briefly  the   phase-field equations for vesicles (or membranes). In section \ref{levelset} we introduce the level-set formulation.   Both formulations  are Eulerian
methods relying on an auxiliary field to capture the interface. We adopt the notation  $\phi$ for the phase-field and $\varphi$ for the level-set function. We then discuss the numerical scheme used for the solution of the  level-set model, and present some numerical results. In section \ref{bridge}  we shall make the bridge between the two methods (level-set and phase-field). We shall show in that section how one can write down a constitutive law for the composite fluid (fluid+membrane). Section \ref{conclusion} is devoted to conclusion and discussion of  some future research directions. Some technical details are relegated into annexes.

\section{Some preliminaries}
\label{preliminaries}
Both phase-field and level-set methods have the common feature that
they define the interface in an implicit manner; that is the phase-field also adopts the level-set notion. A significant difference
exists, however. In a phase-field approach the phase-field values in
the two fluids are forced to remain parallel (or have constant values away from the interface) by imposing a double
well potential, while in a level-set approach the (color) function is
just advected by the flow. Both methods have therefore potential advantages and drawbacks.
Enforcing the phase-field function to given values and controlling the interface width by appropriate potential brings a very stable interface description. However this might introduce volume loss and
 additional terms in the potential are usually added to control this leak. The philosophy of phase-field in this context is therefore to give an Eulerian description of the sharp interface by a diffuse interface of constant and controllable width. The phase-field is a function $\phi$ which is only used to give geometrical information on the interface. By contrast,  in the level-set method  the auxiliary field $\varphi$ is merely advected by the velocity field of the continuous medium, starting initially from a signed distance to the interface. There is no potential to keep the field stuck to a distance function (further details will be given below), and as the diffuse interface
is typically considered to lie between  two given values of $\varphi$, e.g. $-\eps$ and $+\eps$, the interface width varies. The gain to move the level-set function by advection is that some mechanical information is recorded in $\varphi$. For example $\ngvp$ records the area change of interface. This makes it therefore possible to write elastic energy merely in terms of the auxiliary field and to obtain an attractive complex fluid model of the fluid-structure coupling. The non-constant width of interface has however to be taken care of. But this is made as post-processing each time the level-set function is needed to compute the distance to interface.

\section{Phase-field formulation}
\label{phasefield}
In this section let us recall the main results of the phase-field
model introduced earlier\cite{Biben2003,Biben2005}. In that model the
vesicle is described by an advected field $\phi$ which goes smoothly
from $-1$ to $+1$ while crossing the membrane. Let $-1$ represent
the interior of the vesicle and $+1$ its exterior, the membrane
being localized by the zero level-set of $\phi$. In order to ensure
a constant width for the interface, one considers the following
functional
\begin{equation}
E_{intrinsic}[\phi]=\iint\left\{\frac14(1-\phi^2)^2+\frac{\eps^2}2\ngp^2\right\}dx \label{intrinsic}\end{equation}
when minimized in the special case of a flat interface leads to the interfacial profile given by:
$$\phi(r)=\tanh\left(\frac{r}{\eps\sqrt{2}}\right)$$
The above energy is called {intrinsic} in the sense that its role is only to define the boundary, and it should not affect the physics (like the forces acting on the membrane).
The physical forces are accounted for by the introduction of the configurational energy which is given by
\begin{equation}
E_{config}=\frac{\kappa}{2}\iint c(\phi)^2\frac{\ngp}{2}dx+\iint\xi\frac{\ngp}{2}dx\label{energy}
\end{equation}
In this expression the first term stands for the curvature energy with modulus $\kappa$ (which is a phase-field expression of the Helfrich energy), while the second one is a penalization of the membrane {\it local} length variation, $\xi$ being therefore a local (i.e. it depends on the given position and on time) Lagrange multiplier (we do not use  $\zeta$ as in \cite{Biben2003,Biben2005}, since this symbol will denote  a cut-off function thereafter). It is classical that the normal to the interface and projector on its tangent plane are given by
$$n=\frac{\gp}{\ngp},\qquad P=\I-n\otimes n=\I-\frac{\gp\otimes\gp}{\ngp^2},$$
where  ${\I}$ is the identity tensor, while the curvature field is expressed as $c=-\div n$.
Differentiating the above energy with respect to the $\phi$ gives an explicit expression of the external force
\begin{equation}
F_{config}=\left[-\kappa\left\{\frac{c}{2}(c^2-4G)+(P\nabla)^2 c\right\}n+\xi c n+P\nabla\xi \right]\frac{\ngp}{2}\label{fconfig}
\end{equation}
where $G=\det((t_1\cdot\nabla)n,(t_2\cdot\nabla)n,n)$, is the Gaussian curvature and $(t_1,t_2,n)$ an orthonormal trihedron. The non-stationnary Stokes equations are used to find the velocity field:
\begin{equation}
\eps_uu_t-\div\left[\eta(\nabla u+\nabla u^t)\right]-\nabla p=F_{config},\qquad \div u=0\label{Stokes}
\end{equation}
where $\eps_u$ is a relaxation parameter (taken small enough to mimic the zero Reynolds number limit) and $\eta=\eta_{out}(1+\phi)/2+\eta_{in}(1-\phi)/2$ the smeared viscosity (that accounts for a viscosity contrast between the interior and exterior of the vesicle; $\eta_{in}$ and $\eta_{out}$ are the viscosities of the internal and external fluids).
As the membrane is simply advected by flow, a simple transport equation with velocity $u$ (i.e. $d\phi/dt=0$ where $d/dt$ is the material derivative) should be used to find $\phi$. For stability reasons, and to guarantee that the advected field function minimizes the intrinsic energy, and thus it represents the interface in the course of time, the phase-field is taken to obey the equation $d\phi/dt=-\eps_\phi\delta E_{intrinsic}/
\delta \phi$, where $\delta/\delta \phi$ is the functional derivative and $\eps_\phi$ a kinetic constant fixing the time scale. Using  (\ref{intrinsic}) one obtains the following equation:
\begin{equation}
\phi_t+u\cdot\nabla\phi=\eps_\phi(\phi(1-\phi^2)+\eps^2(\Delta\phi+c\ngp))\label{eqphi}
\end{equation}
where the term $c\eps^2\ngp$ has been added by "hand" in order to cancel the wall free energy of the membrane associated with ${\eps^2}2\ngp^2$ (which is known to lead to a surface tension-like term in the force). Since the membrane does not exchange matter with its surrounding environment (unlike drops where molecules from bulk can migrate to surface and vice versa, leading to surface variation), its surface energy is zero. The added terms guarantees the absence of a surface tension \cite{Biben2003,Biben2005} (recently another method to deal with this problem has been suggested \cite{Jamet2008}).

Since the membrane is locally incompressible, one has to impose that
the surface  projected divergence of the velocity field must be
zero. In the sharp interface picture \cite{Cantat1999,Cantat2003} the
local area incompressibility can be handled by introducing a space
and time dependent Lagrange multiplier. This amounts to writing the
contribution of the membrane energy related to the local membrane
incompressibility condition as
\begin{equation}
E_{inc}=\int \zeta ({r_m}, t) dA
\end{equation}
where the integration is performed over the vesicle membrane, and
${r_m}$ is the vector position of the membrane. The Lagrange
multiplier $\zeta$ is then determined by imposing
\begin{equation}
({\I}-{n\otimes n}):\nabla {u}=0.
\end{equation}
The
above expression is nothing but the divergence along the membrane (note that ${\I}-{n\otimes n}$ is the projector).
The Lagrange multiplier $\xi$ does not appear in the above equation
which is the associated constraint, just like the pressure does not
appear in the solenoidal constraint in incompressible hydrodynamics.
However, like the pressure, we show in the following that $\xi$
appears in the other equations of the model (it couples to the
velocity field). In the phase-field spirit
\cite{Biben2003,Biben2005} the idea is to define $\xi$ everywhere
(and not only along the membrane) but confine its action to the
membrane region. We then write
\begin{equation}
E_{inc}=\int \xi (\bfs{r}, t) |\nabla\phi|dx \label{eq:Einc}
\end{equation}
where $dx$ is the volume element of the total domain. Because
$\phi (r)\sim \tanh(r/\sqrt{2} \epsilon) $, it is clear that
$|\nabla\phi|$ is a Dirac-like function of width
$\epsilon$. This implies that the energy acts in the membrane
region only, as it should be.

In \cite{Biben2003,Biben2005} the tension field was postulated to
obey the following equation (apart from the Laplacian term which was
included in \cite{Biben2005} for some numerical regularization)
\begin{equation}
{d\xi\over dt}\equiv \xi_t+u\cdot\nabla\xi=
T{({\I}-{n\otimes n}):\nabla {u}} \label{eqxi}
\end{equation}
$T$ is a tension-like parameter which is
chosen large enough, so that we expect this to enforce
$({\I}-{n\otimes n}):\nabla {u}\simeq 0$, that is a local membrane
quasi-incompressibility. This field was built on the basis of
intuition and knowledge of the sharp interface problem
\cite{Cantat1999,Cantat2003}. It was shown in Ref. \cite{Biben2005} by asymptotic techniques that in the limit $\epsilon\rightarrow 0$ the above equation (\ref{eqxi}) recovers the sharp interface limit (free surface divergence of the velocity field). In \cite{Biben2003,Biben2005} $\xi$ is said to be
proportional to the local extension of the membrane, and indeed the right hand side of (\ref{eqxi}) is a measure of the surface variation, as we shall see. It will be
shown in the next section how equation (\ref{eqxi}) can be  derived.

\section{Level-set formulation}
\label{levelset}
\subsection{Introduction}
Let us now introduce the level-set approach. In the level-set
formulation \cite{osherbook}, one still introduces a function $\vp$
which is negative inside the membrane and positive outside. This
function is not constrained to take values between $-1$ and $+1$.
Rather, it is initially the signed distance to the interface, and is
then advected by a transport equation
\begin{equation}
\vp_t+u\cdot\nabla\vp=0\label{eqvp}
\end{equation}
It was shown in \cite{CotMai,CotMai2} that when $u$ is divergence free, $\ngvp$ records the stretching of the interface $\{\vp=0\}$; this opened the way to a complete formulation in terms of $\vp$ of the membrane forces.
 In order to localize the interface, one introduces a cut-off function $\zeta$, i.e. a non-negative function with compact support included into $[-1,1]$, of unit mass such that $\frac1\eps\zeta(\frac{r}\eps)$ converges to the Dirac mass when $\eps$ goes to zero. We used in our simulations the following expression
 $\zeta(r)=\frac12(1+\cos(\pi r))$ on $[-1,1]$, $\zeta(r)=0$ elsewhere. This function is of unit mass. Considering $\zeta(\frac r\eps)$, this gives a function with support in $[-\eps,\eps]$ which has a mass of $\eps$ as readily seen by integration. This is why the scaling by $\frac1\eps$ is necessary to maintain a unit mass and ensure the convergence of $\frac1\eps\zeta(\frac{r}\eps)$ to the Dirac mass. By composition with the level-set function, we obtain $\ze$ whose support is localized in the strip $-\eps<\vp<\eps$. If $\vp$ is a distance function $\ze$ converges to the Dirac mass on the curve $\vp=0$ when $\eps\to0$. In general  $\ze\ngvp$ converges to this Dirac mass as $\eps\to0$.

Let us first consider the case where the membrane energy would depend on $\ngvp$ only (we shall see that $\ngvp$ is a measure of local surface variation). We shall call the corresponding energy {\it elastic energy} (in contrast to curvature or bending energy). A vesicle membrane  is inextensible. In our spirit we shall allow for a finite, albeit quite small, area variation. If we have in mind a capsule, for example (see review \cite{DBB2009}), then the membrane can undergo a certain local area variation. Therefore our energy may be used to various cases.
We define the energy in the entire space provided that the density energy is multiplied by the appropriate localized function introduced above. We then have
\begin{equation}
\mathcal{E}_m(\vp) = \int_\Omega E(\ngvp)\ze d\bx\label{elasLS}
\end{equation}
and differentiating (i.e. taking functional derivative with respect to $\vp$) it along (\ref{eqvp}) leads to the following elastic force:
\begin{equation}
\bF_m=\left\{\bnabla[E'(\ngvp)]-\div\left[E'(\ngvp)\frac\gvp\ngvp\right]\frac\gvp\ngvp\right\}\ngvp\ze\label{mforce}
\end{equation}
Note that $E$ stands for a constitutive law for the membrane. A trivial example is a linear law $E'(r)=\Lambda(r-1)$ \footnote{In \cite{Raoult} it was shown that $E'(r)=\Lambda\max(r-1,0)$, is a more appropriate choice, since it can be obtained from a variational asymptotic derivation}. $\Lambda$ is a parameter. For vesicles $\Lambda$ is taken quite large in order to enforce quasi-inextensibility.

 A prior study \cite{CotMai,CotMai2} has already considered these  membrane elastic forces, but not
 the curvature energy, which we consider here. We set
\begin{equation}
\mathcal{E}_c(\vp)=\int_\Om G(\kp)\ngvp\ze\label{courbLS}
\end{equation}
where $G$ is again a constitutive law for the curvature energy. A standard choice, which is compatible with \cite{Biben2003,Biben2005}, is $G(r)=\frac{\kappa}2 r^2$, which is nothing but the Helfrich curvature energy density. Recall that $c$ is the divergence of the normal vector, and hence $c(\varphi)=\div\frac\gvp\ngvp$ (note that we have the $+$, while in Refs \cite{Biben2003,Biben2005} the opposite was chosen), which is thus  positive for convex vesicles. Our strategy to derive the force from the energy is the following: we compute the time derivative of the  energy and equal it to minus the power of the curvature force (this is sometimes called  the virtual power --note that a direct functional derivative can be used as well, as in Ref.\cite{Biben2005}):
$$\frac{d\;}{dt}\mathcal{E}_c(\vp)=-\int_\Om F_c\cdot \bu d\bx$$
From differential calculus and using the transport equation (\ref{eqvp}) solved by $\vp$, we have
$$\frac{d\;}{dt}\mathcal{E}_c(\vp)=d\mathcal{E}_c(\vp)(\vp_t)=d\mathcal{E}_c(\vp)(-\bu\cdot\gvp)$$
where $d\mathcal{E}_c(\vp)(\delta)$ means the differential of $\mathcal{E}_c$ at point $\vp$ applied to the increment $\delta$. Therefore we only need to compute the differential of curvature energy and apply it to the increment $-\bu\cdot\gvp$. We show in annex I that
$$d\mathcal{E}_c(\vp)(\delta)=\int_\Om\div\left[-G(\kp)\frac\gvp\ngvp+\frac1\ngvp\prpo\left(\nabla[\ngvp G'(\kp)]\right)\right]\ze\delta dx$$
which by identification leads to
\begin{equation}
F_c=\div\left[-G(\kp)\frac\gvp\ngvp+\frac1\ngvp\prpo\left(\nabla[\ngvp G'(\kp)]\right)\right]\ze\gvp\label{cforce}
\end{equation}
The equations for $u$ in the level-set formulation is similar to the phase-field formulation (see (\ref{Stokes})), with $F_m+F_c$ as source term. In fact we rather use the full  Navier-Stokes equations (i.e. by including $u.\nabla u$) but this induces non significant differences as long as the Reynolds number is small enough. Note that in this formulation there is no need to introduce a Lagrange multiplier and to postulate some corresponding evolution equation (as presented in the last section), since the stretching is encoded the $\vp$ variable. Note also that this implies (unlike in the case of the phase-field approach), that we have to solve  a transport equation for $\vp$, since it is under this evolution that $\ngvp$ records the membrane stretching (see next section and section \ref{Link}). Before discussing the bridges between the phase-field and level-set methods, we shall first present the numerical method used for the level-set method along with some numerical results.

\subsection{Numerical procedure}
\label{Numerical}
The level-set model amounts to solve the following set of equations for $(u,\vp)$:
\begin{align}
\rho(u_t+u\cdot \nabla u)-\div(\eta D(u)) +\nabla p&=F_m+F_c\\
\div u &=0\\
\vp_t+u\cdot\nabla \vp&=0
\end{align}
with appropriate initial and boundary conditions.

As there is no additional term in the transport equation, $\vp$ does not remain a distance function for $t>0$. To see this, we use the fact that distance function are functions with unit gradient modulus. Differentiating (\ref{eqvp}), one easily find that $\ngvp$ verifies
$$\ngvp_t+u\cdot\nabla\ngvp=-\ngvp\frac{\gvp\otimes\gvp}{\ngvp^2}:D(u)$$
therefore, starting from $|\gvp^0|=1$ initially, we have $\ngvp=1$ for $t>0$ if and only if
$$\frac{\gvp\otimes\gvp}{\ngvp^2}:D(u)=0$$
for all time. This means that the variation of $u$ along the direction normal to the interface is zero, which is not true in general.
As a consequence the support of the smeared delta function $\ze$ will vary, making the smeared membrane width locally proportional to the inverse of extension. To circumvent this effect, one uses the following trick: we use $\frac{\vp}{\ngvp}$ as an approximation of the distance function to the interface, thus replacing $\ngvp\ze$ by $\frac1\eta\zeta(\frac{\vp}{\ngvp\eta})$. This order one approximation could seem brutal, but several numerical evidences \cite{CotMai2} proved its efficiency  and precision in comparison with the usual renormalization trick used in level-set methods \cite{osherbook}.

The numerical scheme used is a Chorin projection method on a MAC mesh, which enforces the exact divergence free condition, and therefore the volume conservation, at the discrete level. This is of high importance for our problem since change in the volume affect the shape of the minimizer of the curvature energy.

\subsection{Dimensionless parameters} The level-set main equation is given by

\begin{multline*}
\rho(u_t+u\cdot \nabla u)-\div(\eta D(u)) +\nabla p=\left\{\bnabla[E'(\ngvp)]-\div\left[E'(\ngvp)\frac\gvp\ngvp\right]\frac\gvp\ngvp\right.\\\left.+\div\left[-G(\kp)\frac\gvp\ngvp+\frac1\ngvp\prpo\left(\nabla[\ngvp G'(\kp)]\right)\right]\frac{\gvp}{\ngvp}\right\}\ngvp\frac{1}{\eps}\zeta\left(\frac{\vp}{\eps}\right)
\end{multline*}
Let $L$, $U$, $\rr$ and $\mr$ represent  characteristic length, velocity, density and viscosity scales. Accordingly we set  $x=Lx'$, $u=Uu'$, $t=(L/U)t'$, $\rho=\rr\rho'$, $\eta=\mr r$, $p=\mr (U/L) p'$, $\phi = L\phi'$, and $\eps=L\eps'$.  Differentiating we find\begin{multline*}
u_t=\frac{U^2}{L}u'_{t'},\quad \nabla u= \frac{U}{L}\nabla' u',\quad D(u)=\frac{U}{L}D'(u'),\quad\div(\eta D(u))=\frac{U\mr}{L^2}\div'( r D'(u')),\\ \nabla p = \mr \frac{U}{L^2}\nabla' p',\quad \nabla\vp = \nabla\vp', \quad c(\vp)=\frac1L c'(\vp')\end{multline*}
In dimensionless variables (dropping the $'$), and for the particular case $E'(r)=\Lambda(r-1)$ and $G(r)=\frac\kappa 2r^2$ we get:
\begin{multline*}
Re\rho(u_t+u\cdot \nabla u)-\div(r D(u)) +\nabla p=\left\{\frac1{W_e}\left[\bnabla[E'(\ngvp)]-\div\left[E'(\ngvp)\frac\gvp\ngvp\right]\frac\gvp\ngvp\right]\right.\\\left.+\frac1{C_k}\div\left[-G(\kp)\frac\gvp\ngvp+\frac1\ngvp\prpo\left(\nabla[\ngvp G'(\kp)]\right)\right]\frac{\gvp}{\ngvp}\right\}\ngvp\frac{1}{\eps}\zeta\left(\frac{\vp}{\eps}\right)\end{multline*}
where
$$R_e = \frac{LU\rr}{\mr},\qquad W_e = \frac{\mr U}\Lambda,\qquad C_k=\frac{\mr U L^2}{\kappa}$$
are the Reynolds, Weissenberg and capillary numbers.
In a shear flow one important quantity is the shear rate $\gamma$, from which we can express the characteristic velocity $U=\gamma L$. Substituting we finally get
$$W_e = \frac{\mr\gamma L}\Lambda,\qquad C_k=\frac{\mr \gamma L^3}{\kappa}$$
which are the dimensionless parameters from \cite{Biben2003,Biben2005}. $C_k$ is a measure of bending
distortion of the vesicle. The higher $C_k$ is the easiest the bending mode is (note that bending rigidity appears in the denominator. $W_e$ measures the stretching (or dilatation) modes. Thus $\Lambda$ has to be taken large enough in order to prevent significant membrane extensibility.

There are two additional dimensionless parameters, namely the viscosity contrast $\lambda$ and the reduced volume (or reduced area in two dimensions) $\nu$, defined by
\begin{equation}
\lambda={\eta_{in}\over \eta_{out}}, \;\; \nu_{3D}= {6\sqrt{\pi}V\over A^{3/2}},\;\;\nu_{2D}= \frac{A}{\pi[P/2 \pi ]^{2}} \label{nu}
\end{equation}
where $V$ (respectively $A$) is the vesicle volume (respectively area) and $P$ is the perimeter in the two dimensional problem. In three dimensions (respectively two dimensions), $\nu$ corresponds to the ratio between the actual volume (respectively area) over the volume (respectively area) of the a sphere (respectively circle)  having the same area (respectively perimeter). For a sphere (or circle) $\nu=1$, and it is less otherwise. As an example a human RBC has a reduced volume $\nu\simeq 0.65$. The numerical results presented below correspond to two dimensions, so $\nu$ will refer to $\nu_{2D}$ introduced in (\ref{nu}).

\subsection{Numerical results}
\begin{figure}
\includegraphics[width=3.5cm]{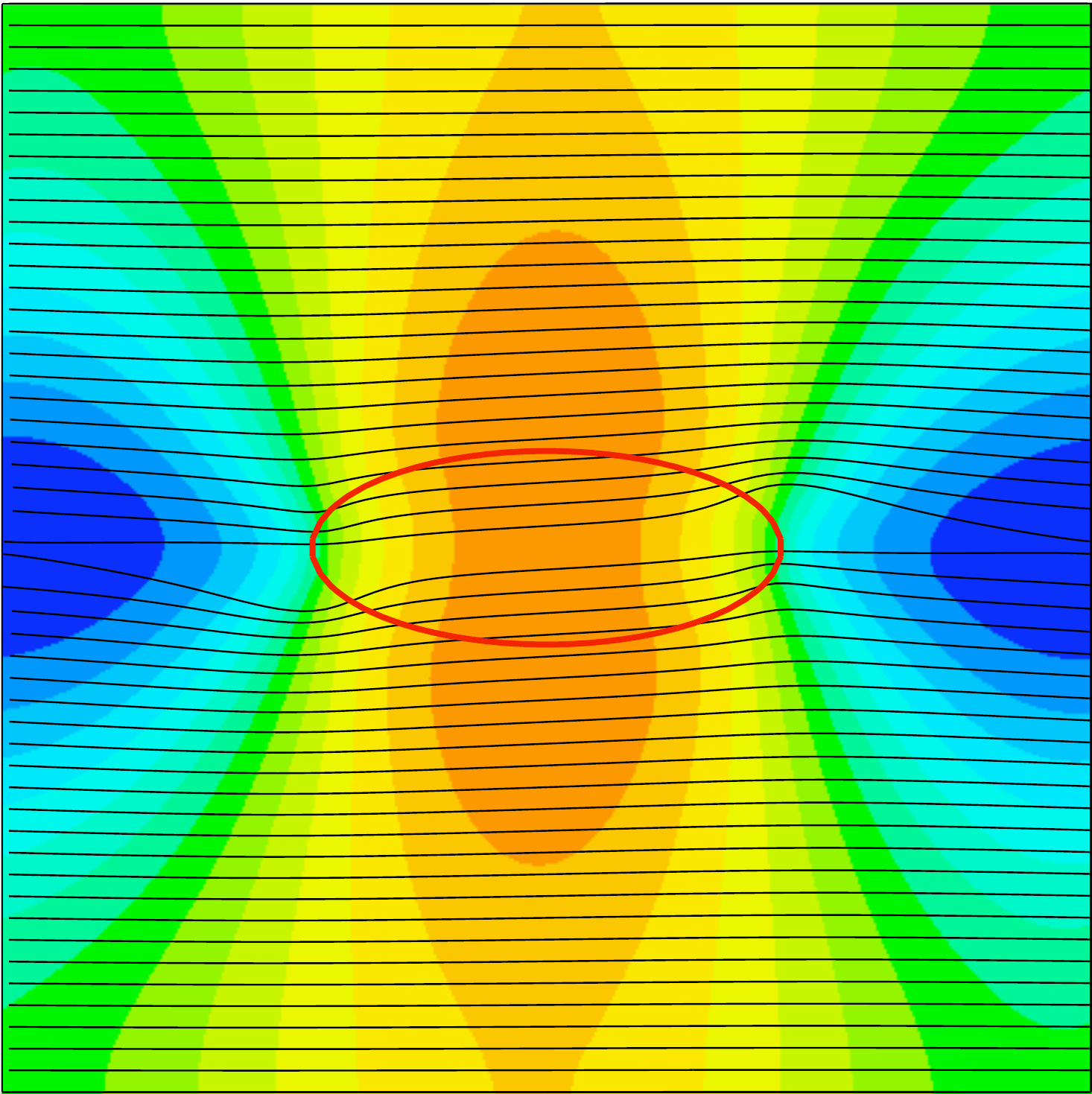}\includegraphics[width=3.5cm]{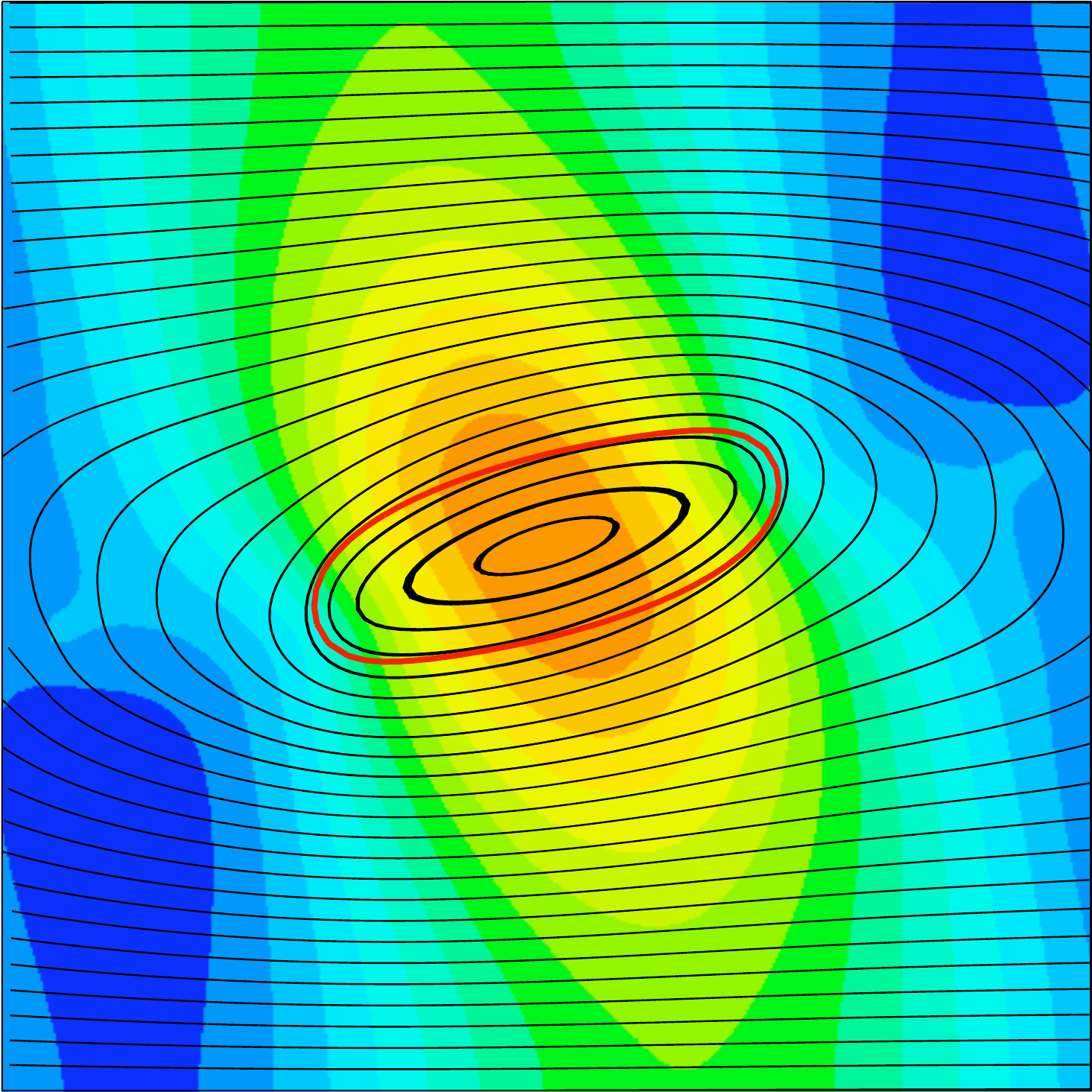}\includegraphics[width=3.5cm]{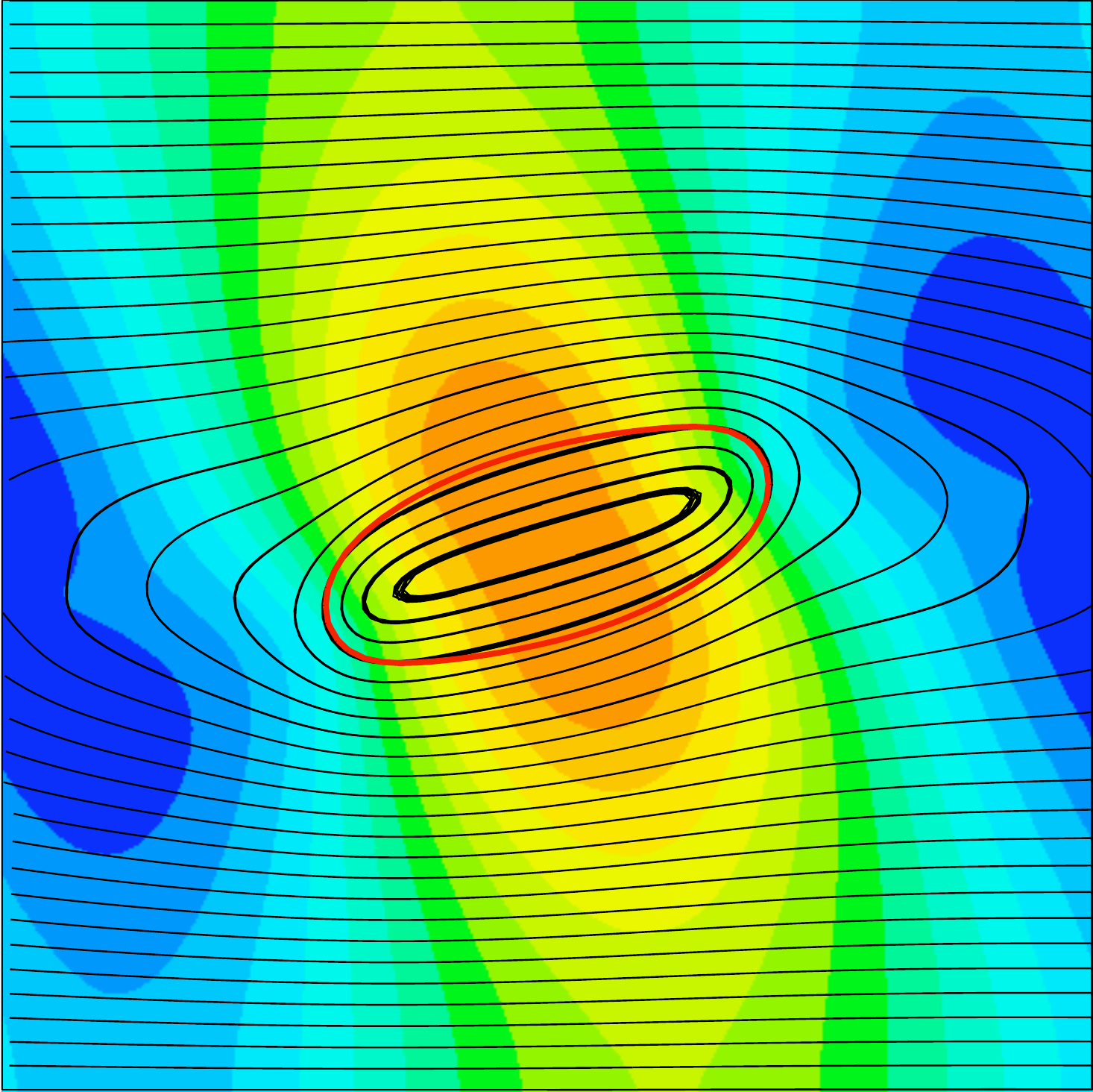}\includegraphics[width=3.5cm]{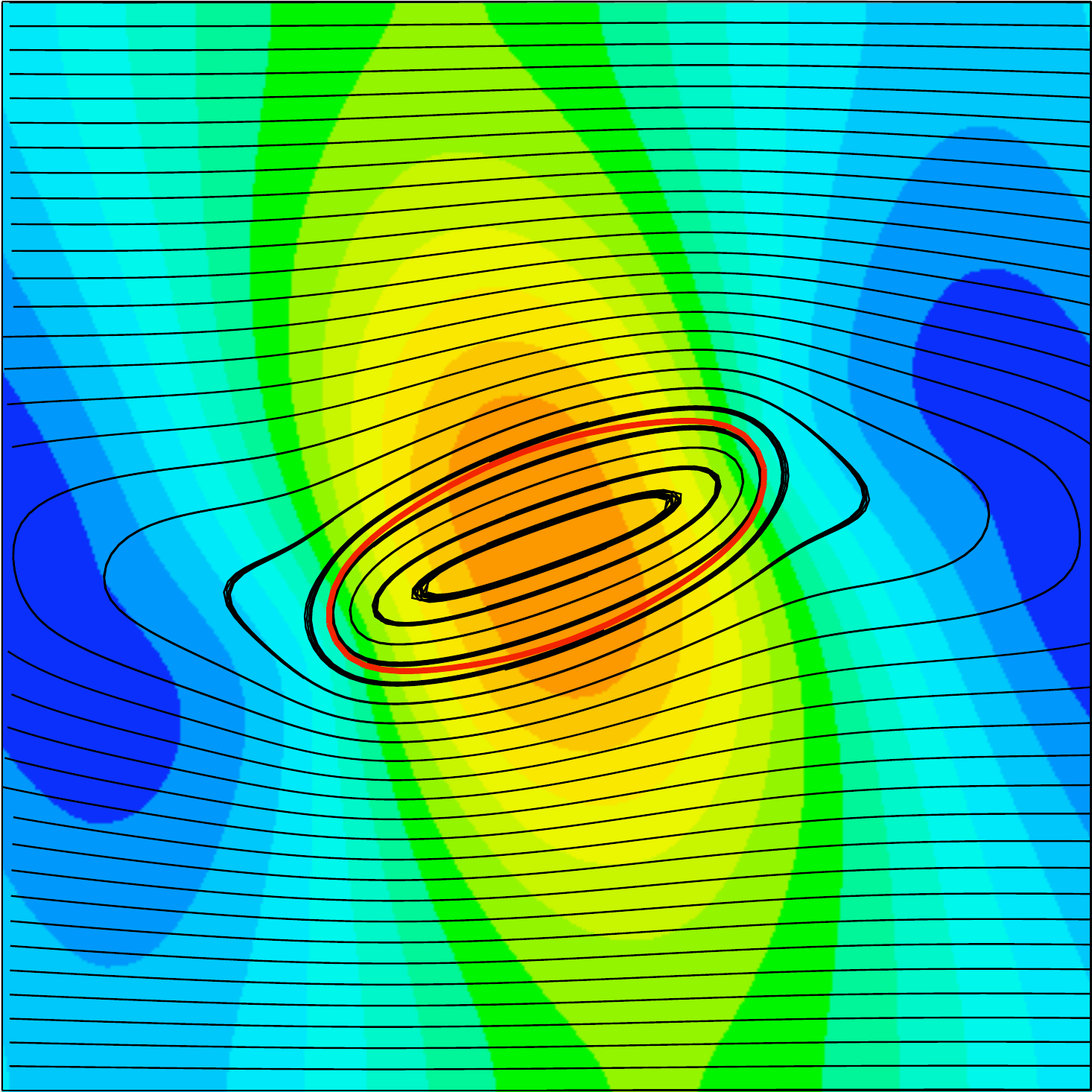}\includegraphics[width=3.5cm]{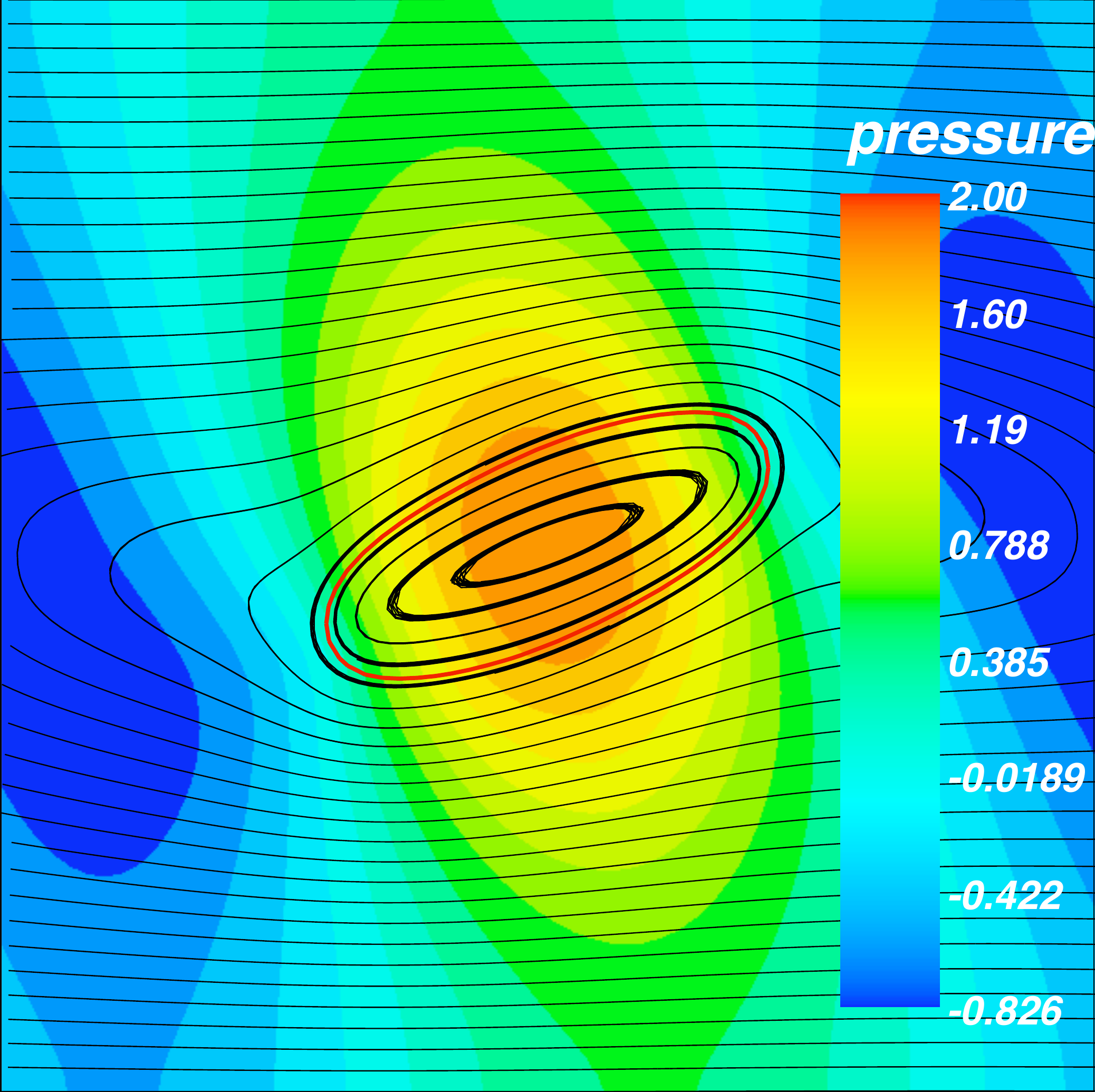}
\caption{\label{tank}Stream lines for the tank-treading motion of a vesicle in a shear flow, at times $10^{-5}$, $5$, $10$, $15$ and $20$. Viscosity contrast: $\lambda=1$, reduced area $\nu=0.7$. Colors stand for iso-pressure lines. }
\end{figure}

Below we present  simulation results corresponding to generic situations where the vesicle makes the classical tank-treading or tumbling motion, and less classical motion called vacillating-breathing  (an intermediate regime which  has been described theoretically by one of us \cite{Misbah2006}, and experimentally by Podgorsky and Steinberg). We set  $Re=0.0001$ (quite close to the Stokes limit), $W_{e}=0.000025$, and $C_{k}=0.25$.
Note that the fact that $C_{k}$ is much larger than $W_e$ means that at the time scale of bending modes (that time is given typically  by ${\mr  L^3}/{\kappa}$) of the vesicle is larger than the time of the elastic mode (that time is typically given by $\mr L\Lambda$. In other words the vesicle elastic response is quasi-instantaneous (on the scale on the physical bending response) and tries to keep the local area (perimeter in two dimensions) as close as possible to the initial one. Typical variations of the observed variations is of the order of percent. The volume is conserved with a much better accuracy (typically $1\%$ for $N=64$ and $0.1\%$ for $N=128$ at the end of the presented simulations). We first fix   the viscosity contrast $\lambda=1$. We observe that the vesicle reaches a stationary angle (\ref{tank}), as is expected. The membrane (which is fluid) undergoes a tank-treading motion.

\begin{figure}
\begin{center}
\includegraphics[width=4.25cm]{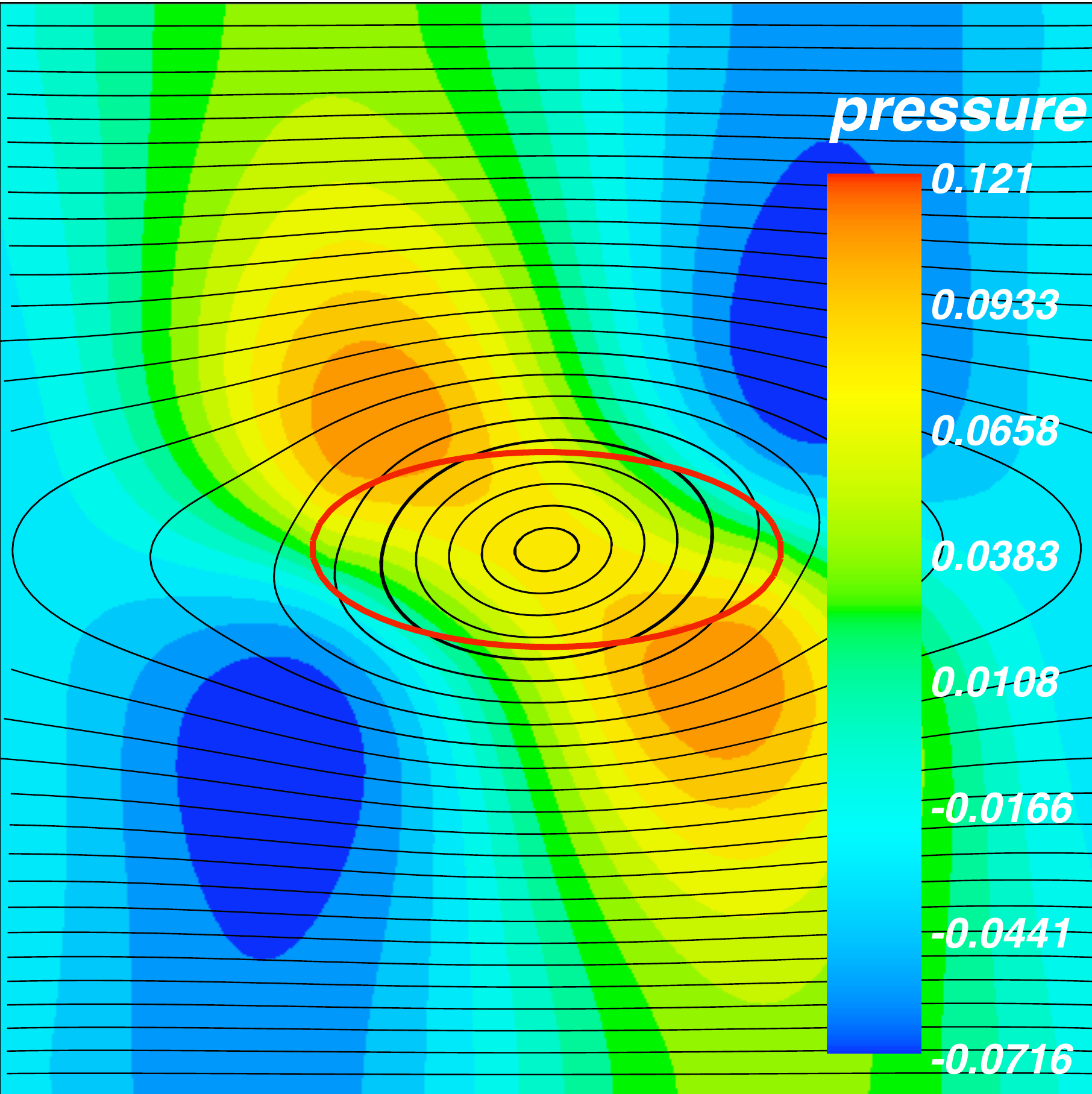}\includegraphics[width=4.25cm]{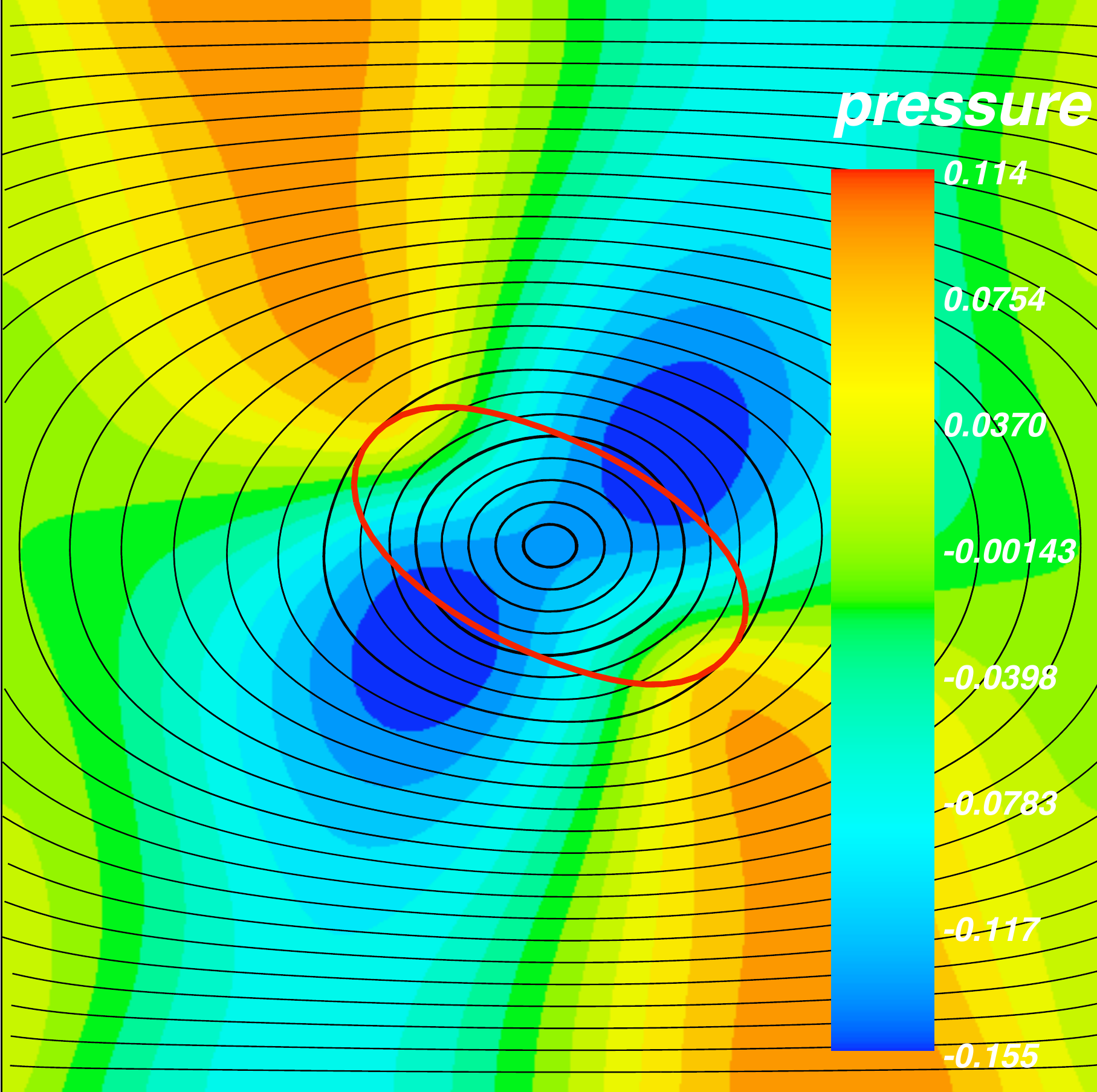}\includegraphics[width=4.25cm]{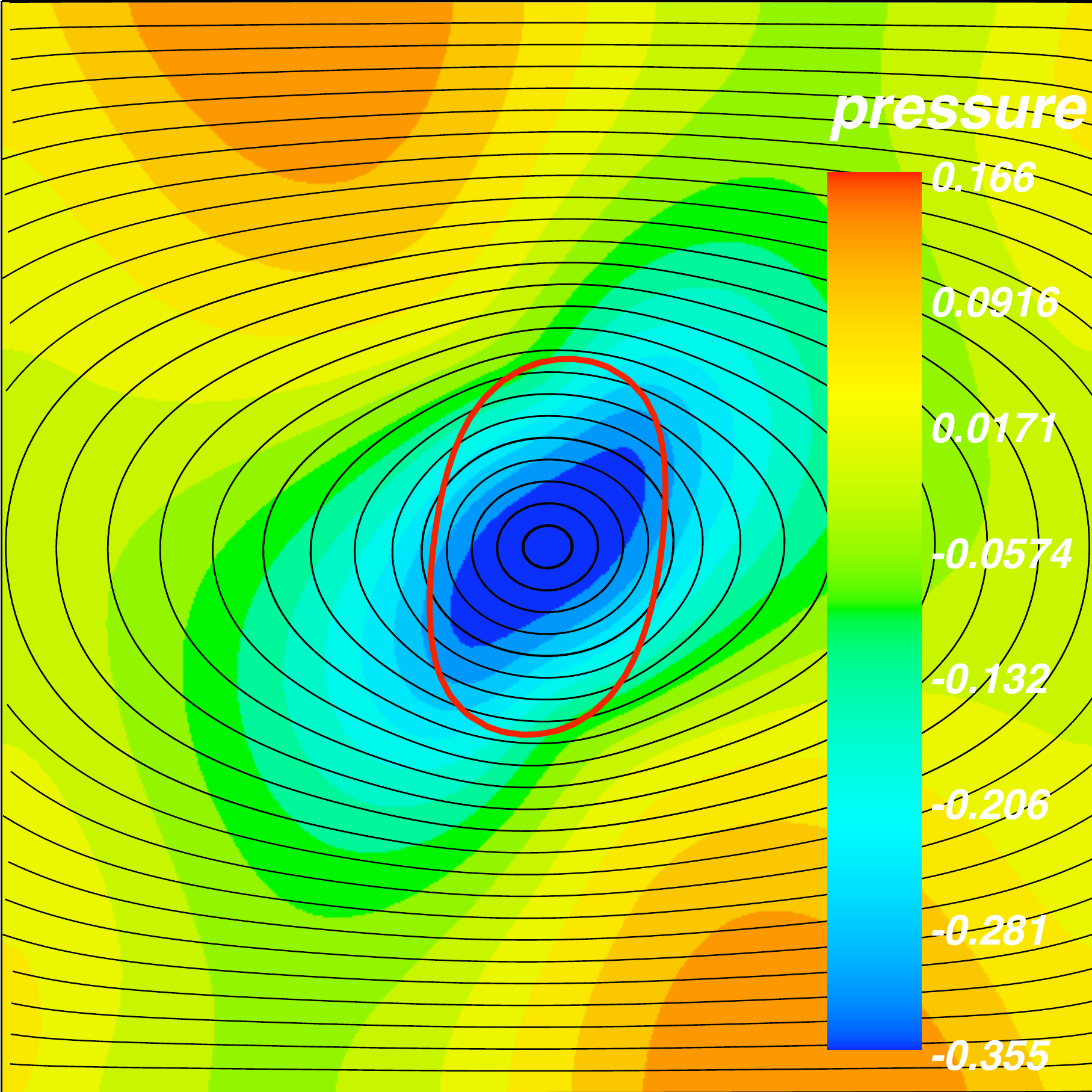}\includegraphics[width=4.25cm]{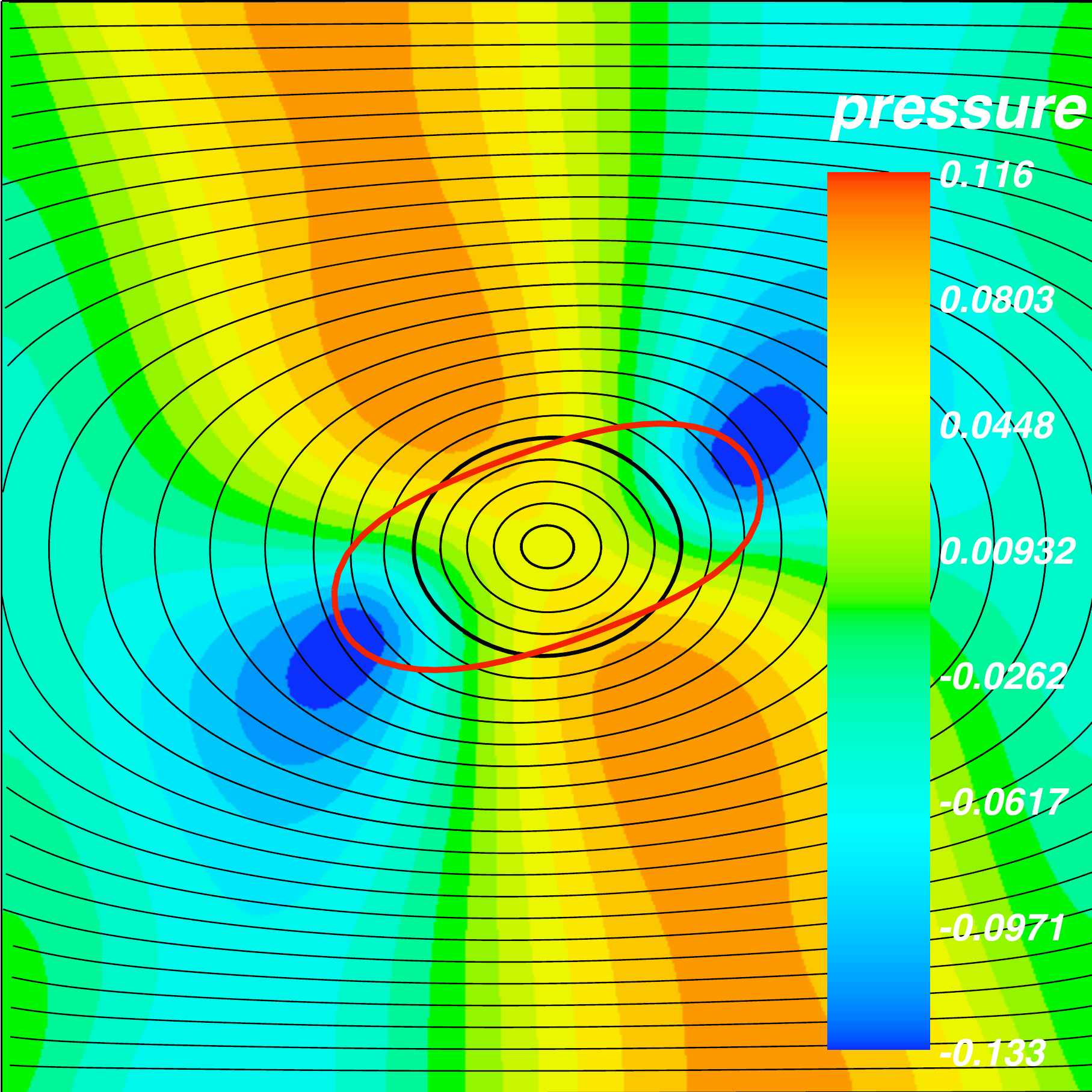}\\
\includegraphics[width=4.25cm]{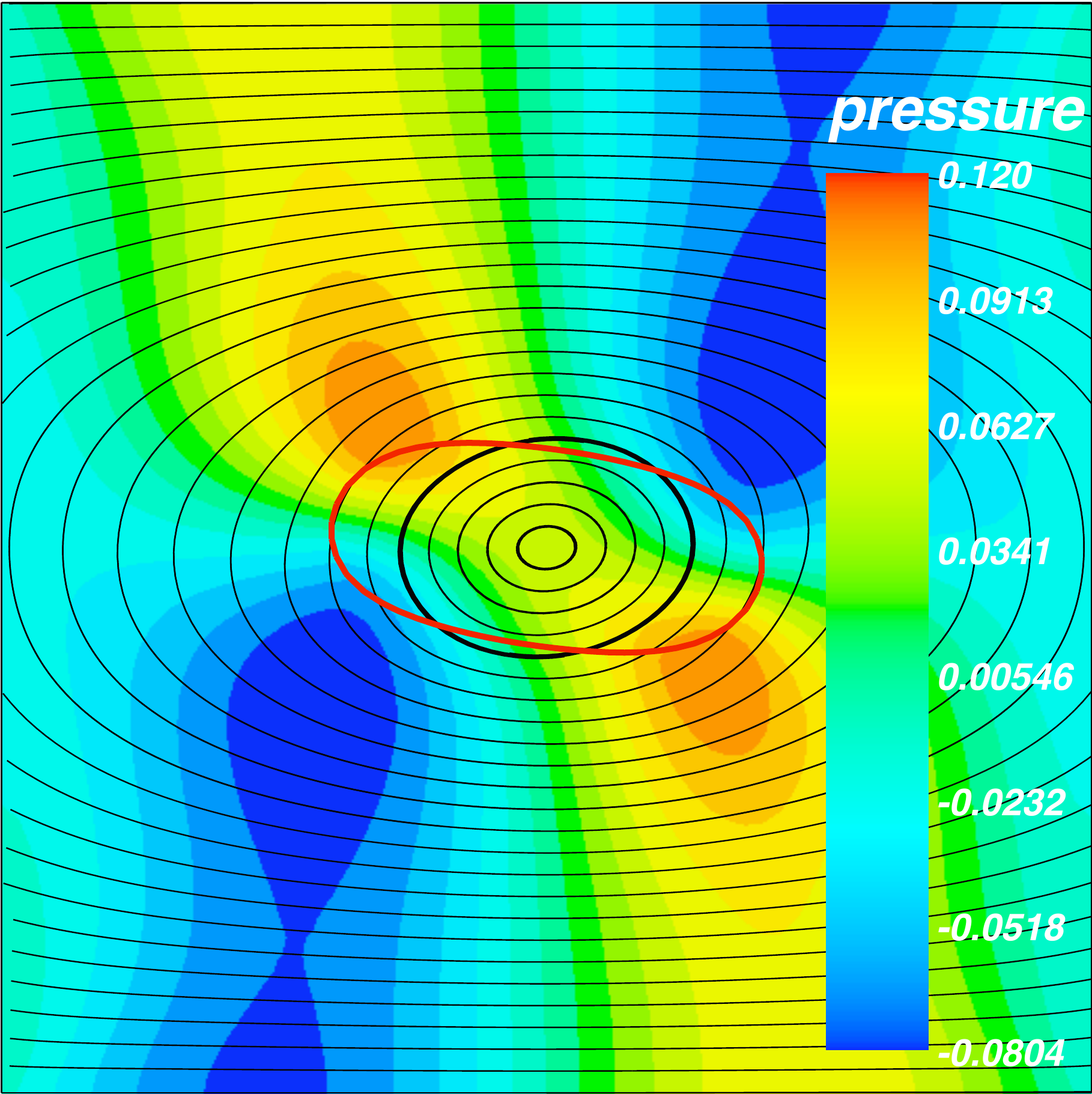}\includegraphics[width=4.25cm]{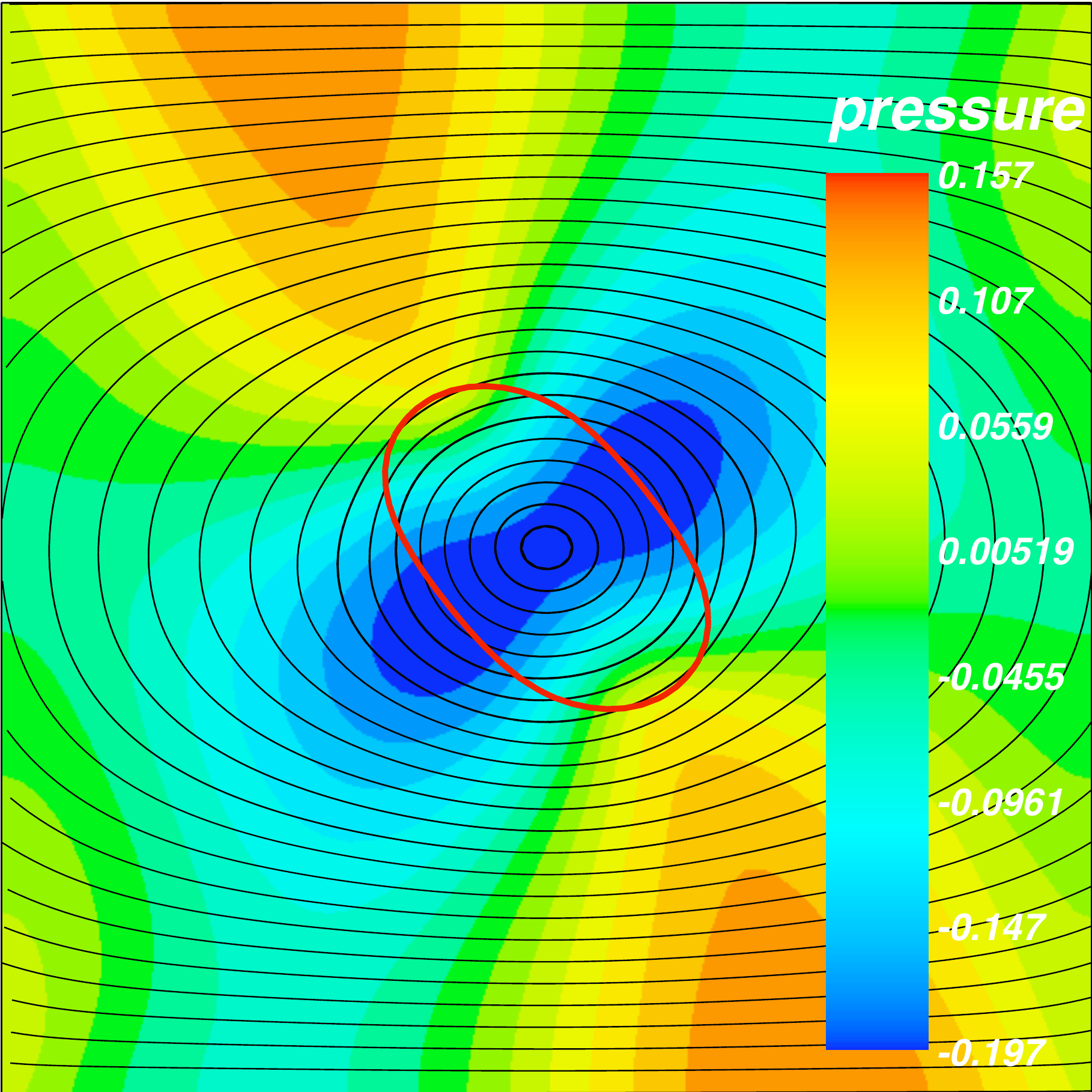}\includegraphics[width=4.25cm]{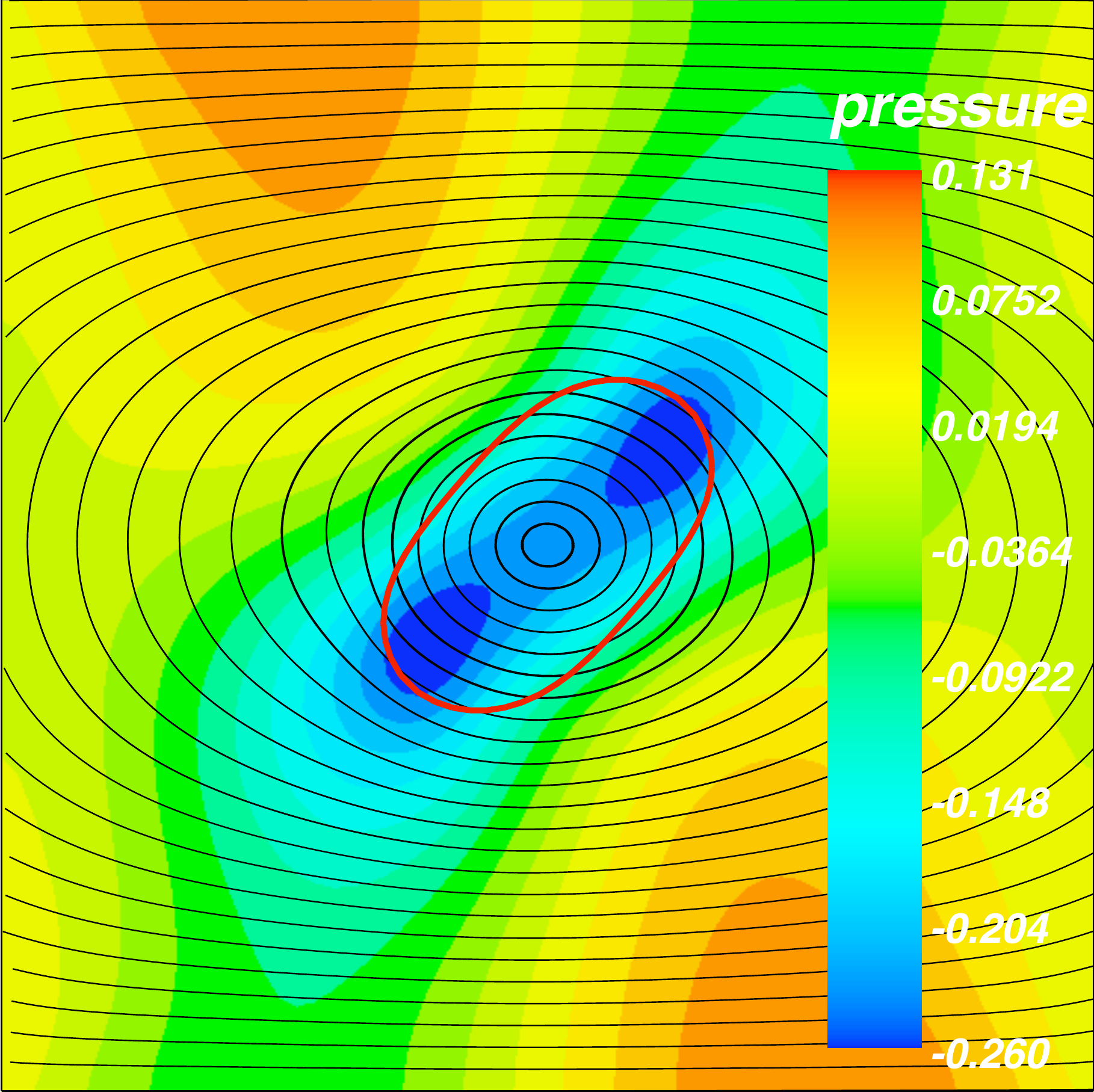}\includegraphics[width=4.25cm]{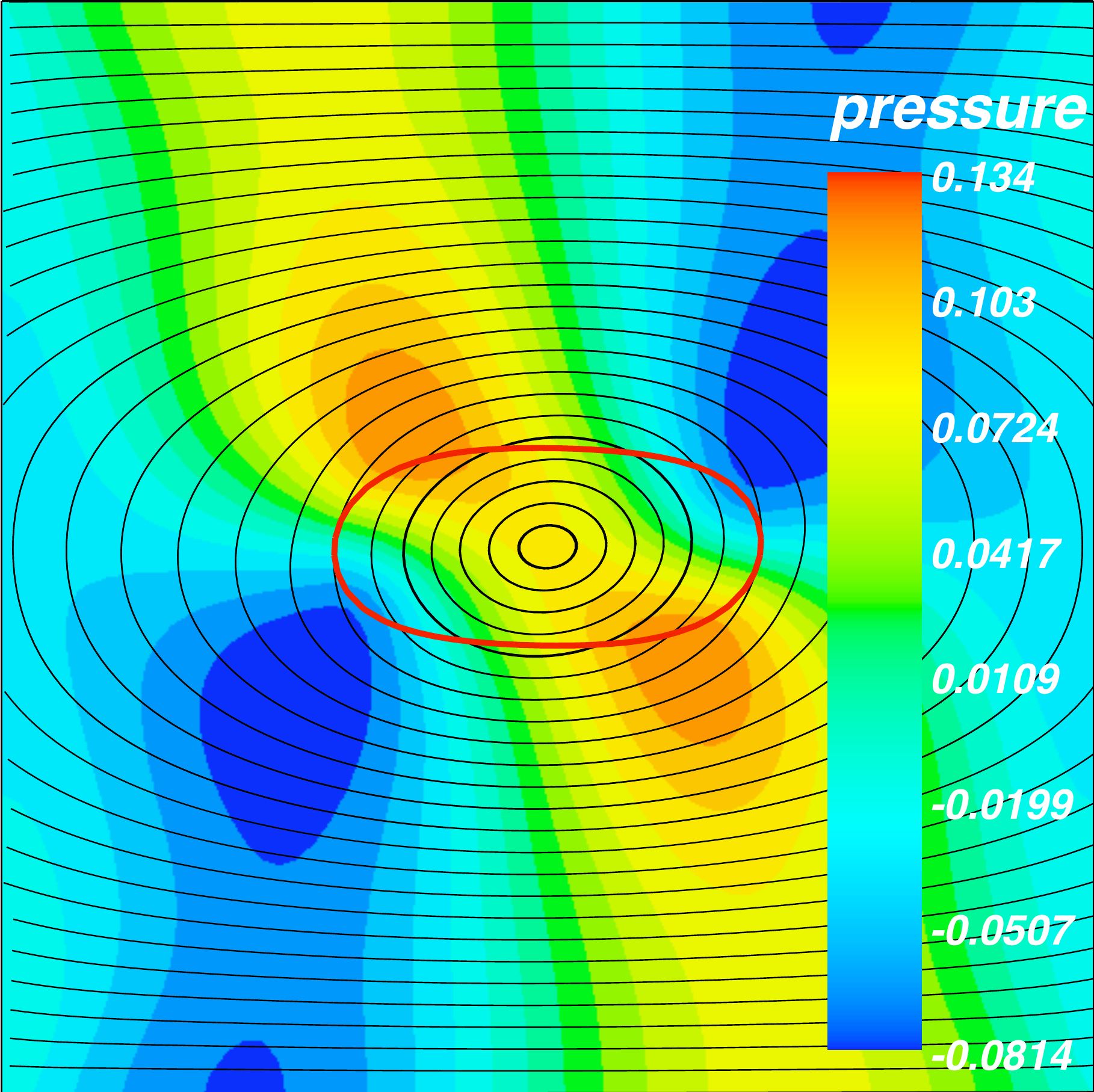}
\caption{\label{tumb} Stream lines for the tumbling motion of vesicle in a shear flow. Viscosity contrast: $8$ at times $0.2$, $4$, $8$, $12$, $16$, $20$, $24$, $29$. Reynolds: $10^{-4}$. Colors stand for iso-pressure lines.}
\end{center}
\end{figure}

\begin{figure}
\begin{center}
\hspace*{-2mm}\includegraphics[width=3.5cm]{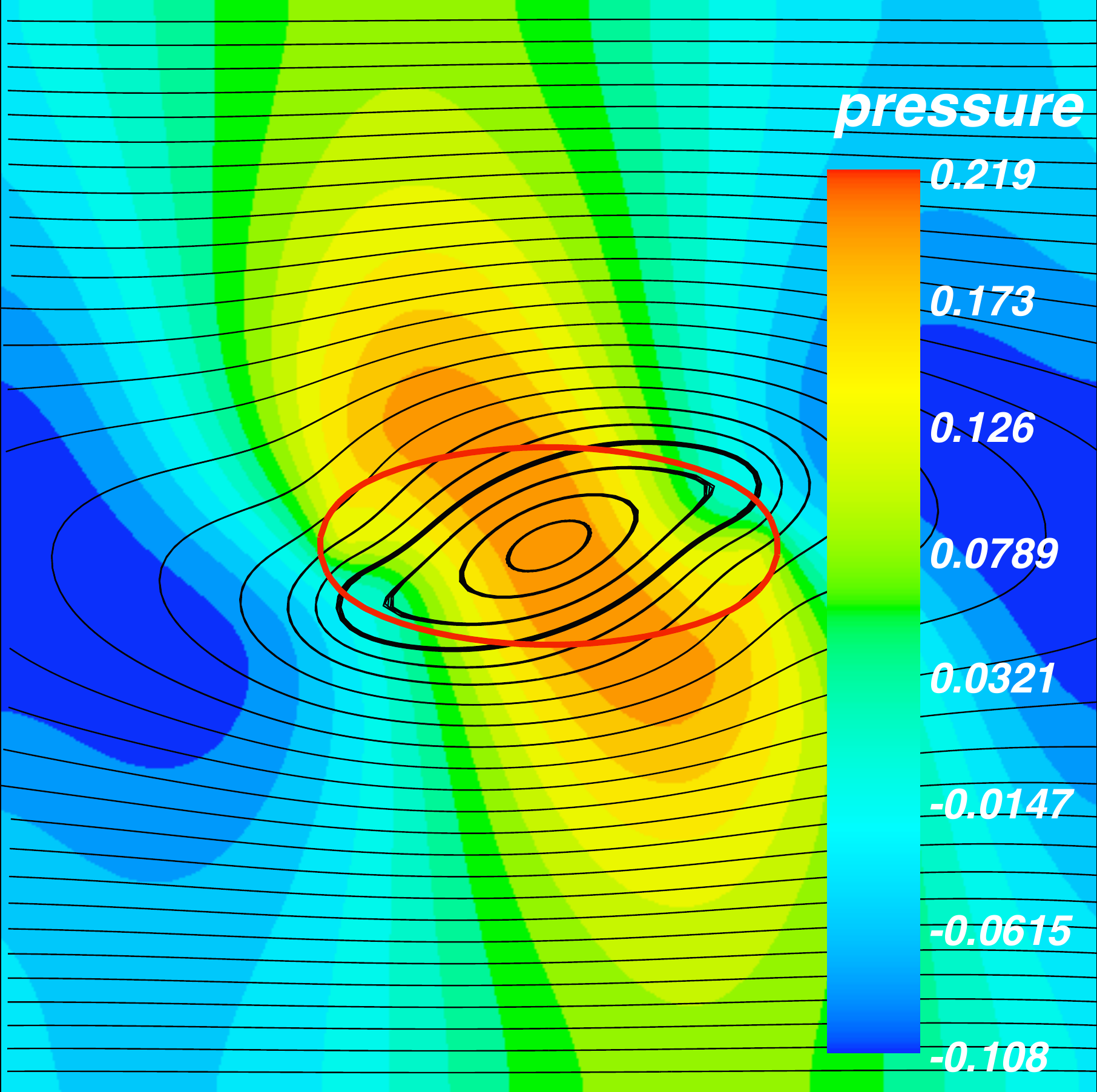}\includegraphics[width=3.5cm]{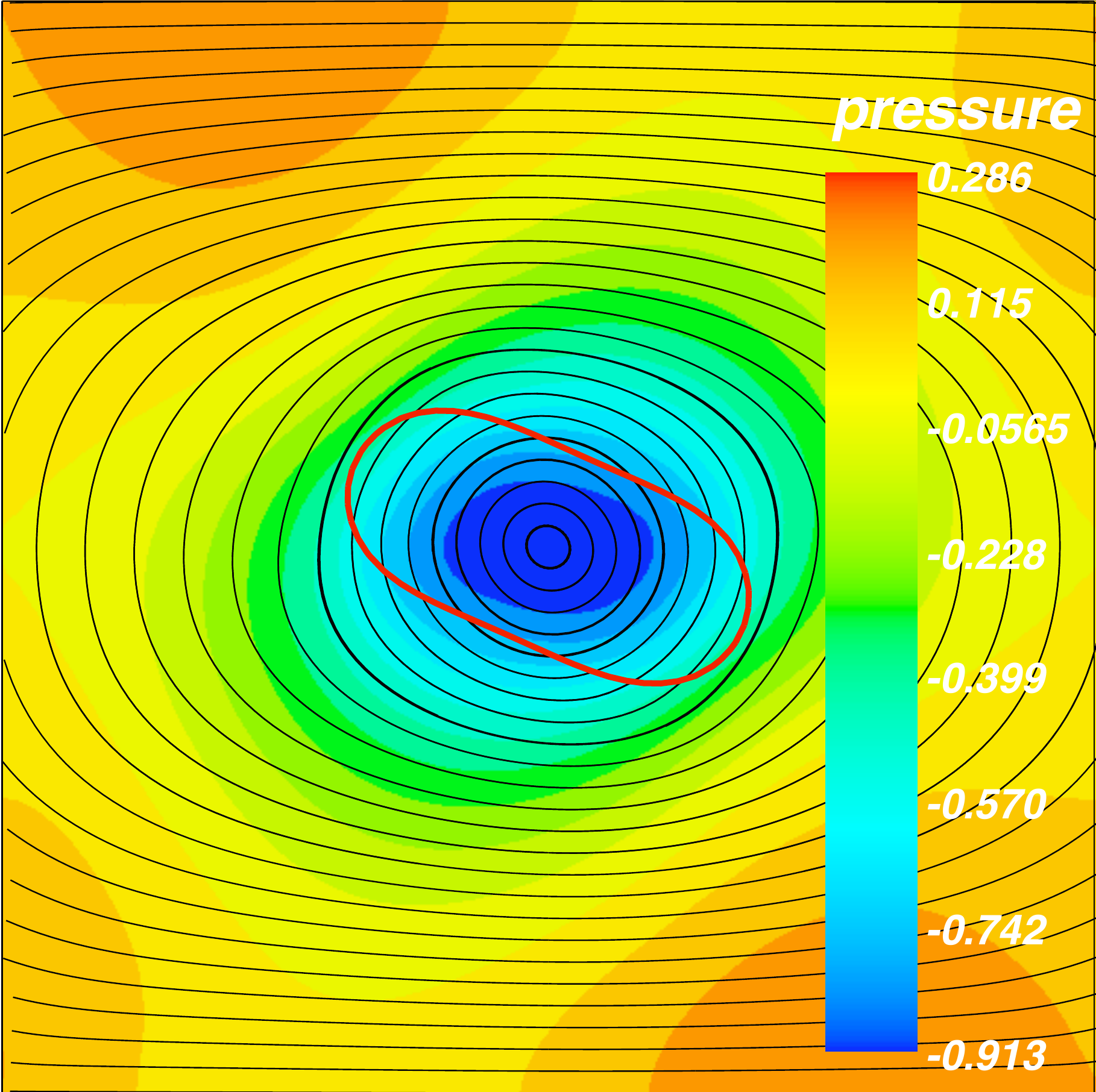}\includegraphics[width=3.5cm]{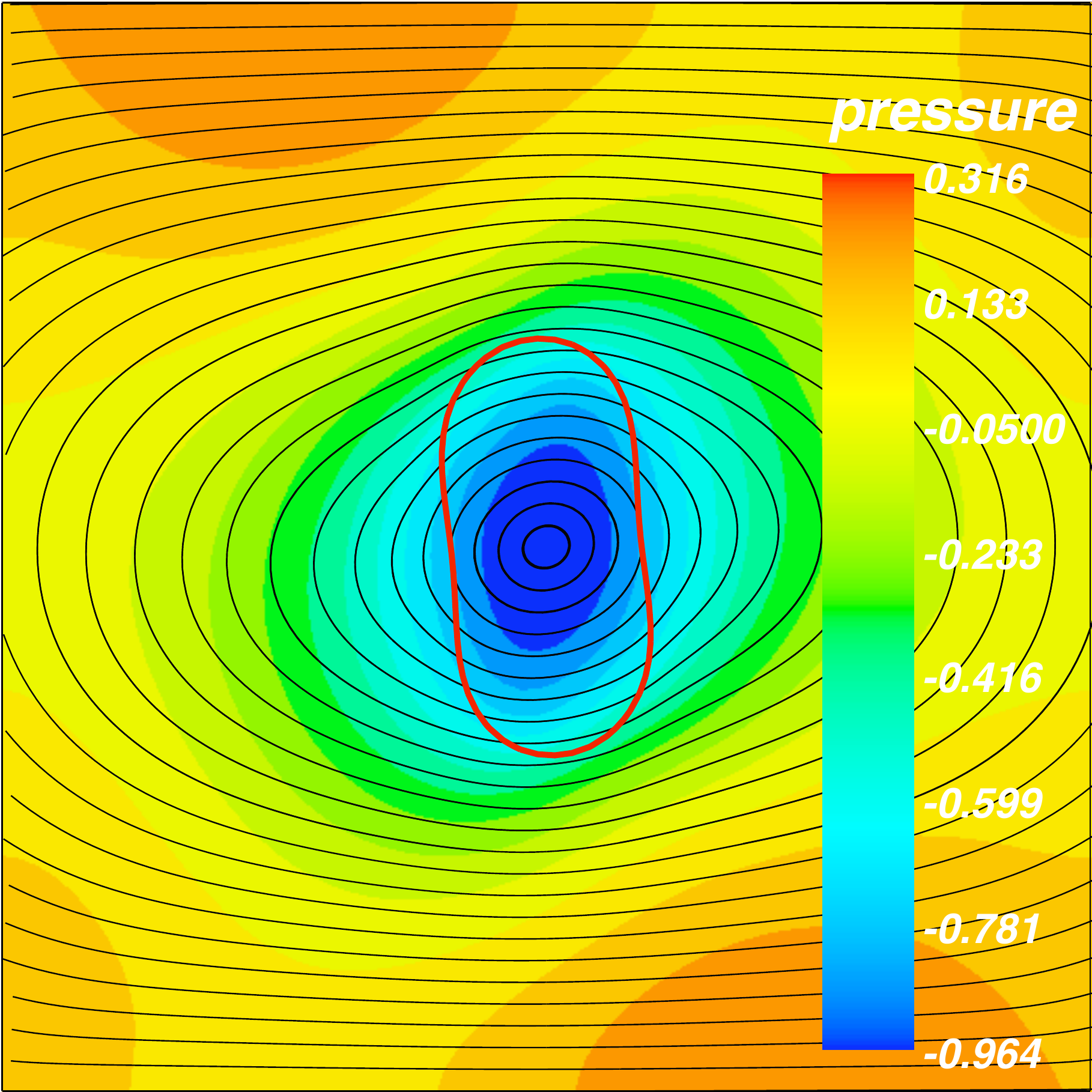}\includegraphics[width=3.5cm]{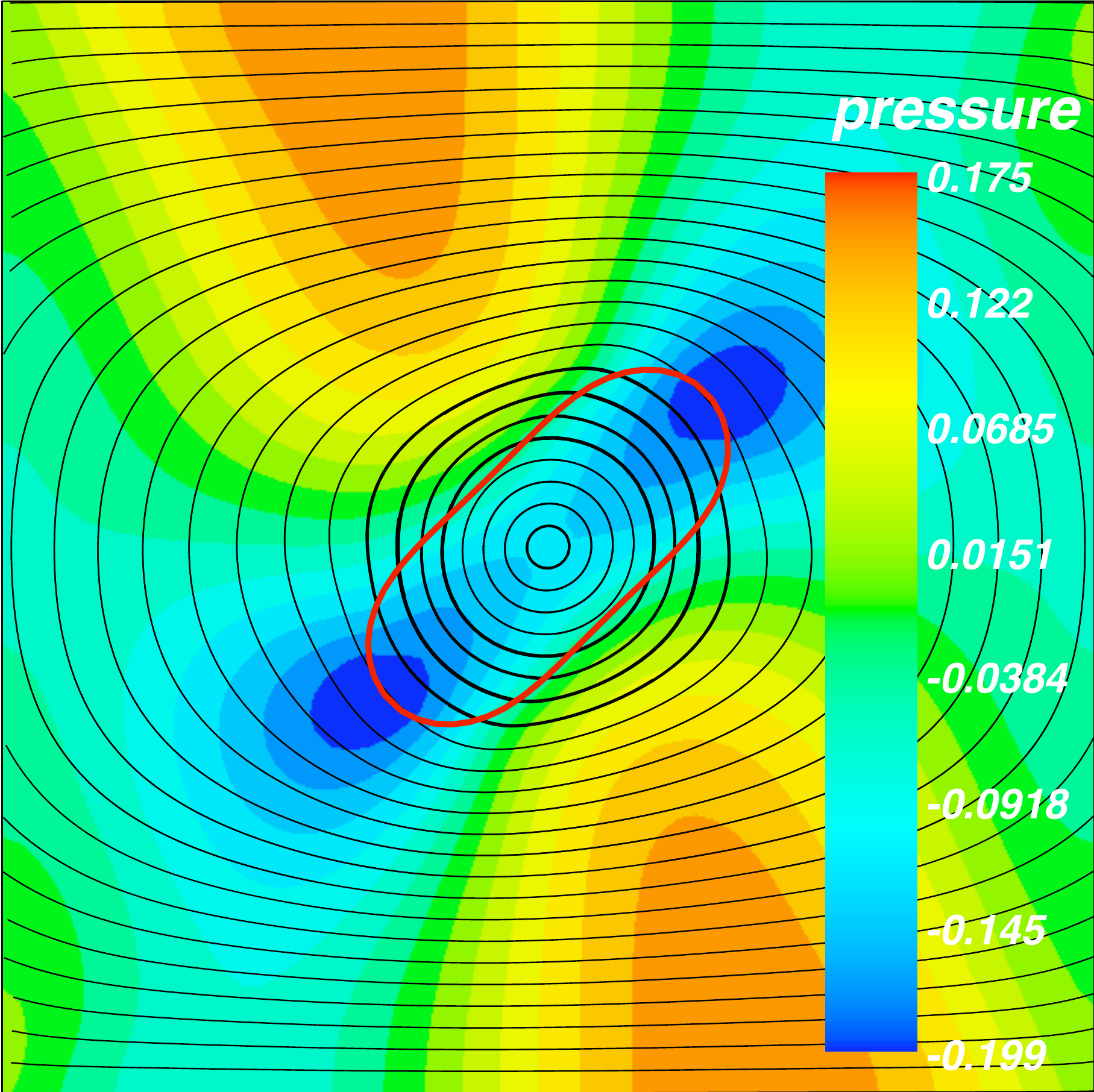}\includegraphics[width=3.5cm]{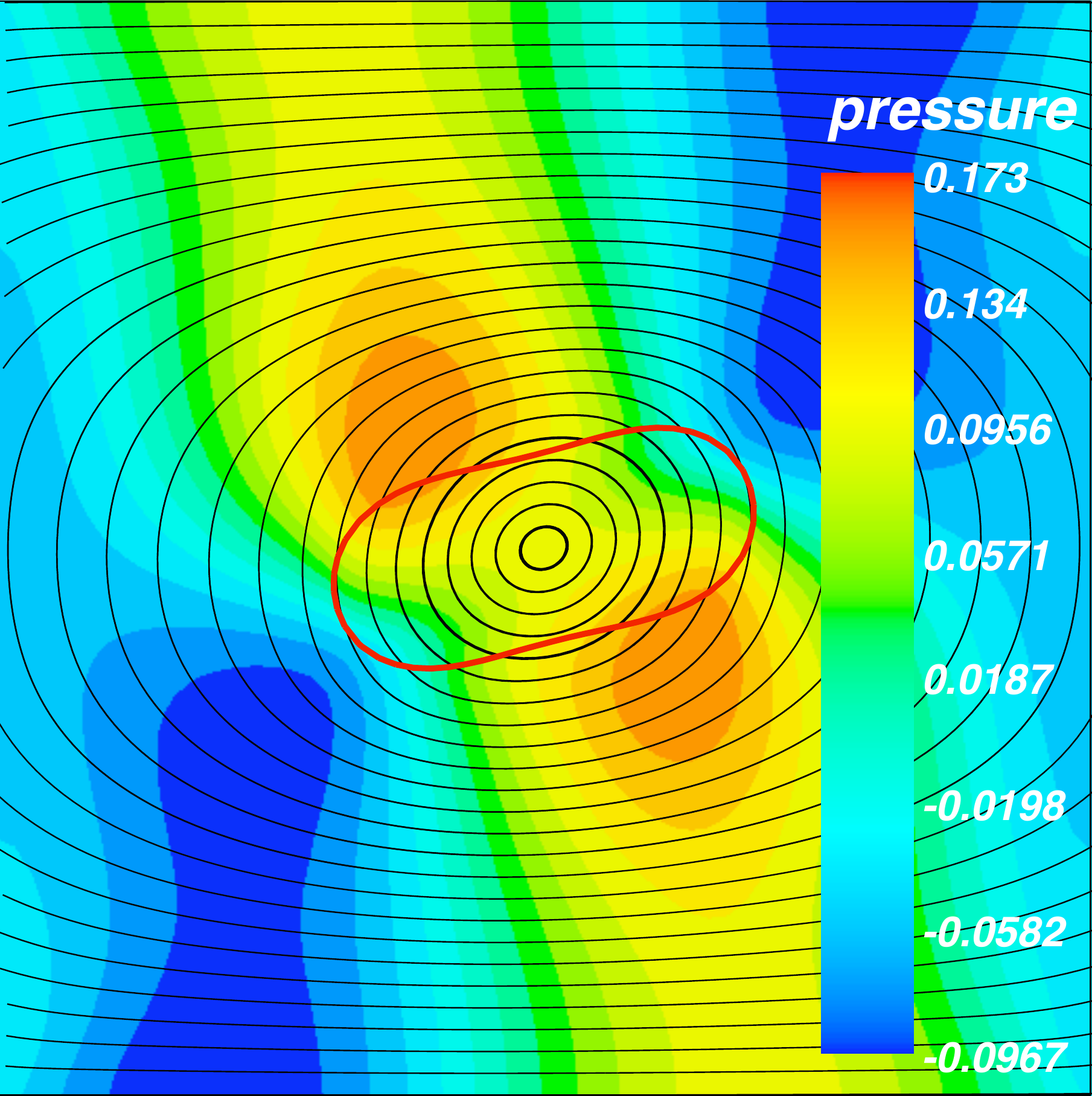}\\
\hspace*{-2mm}\includegraphics[width=3.5cm]{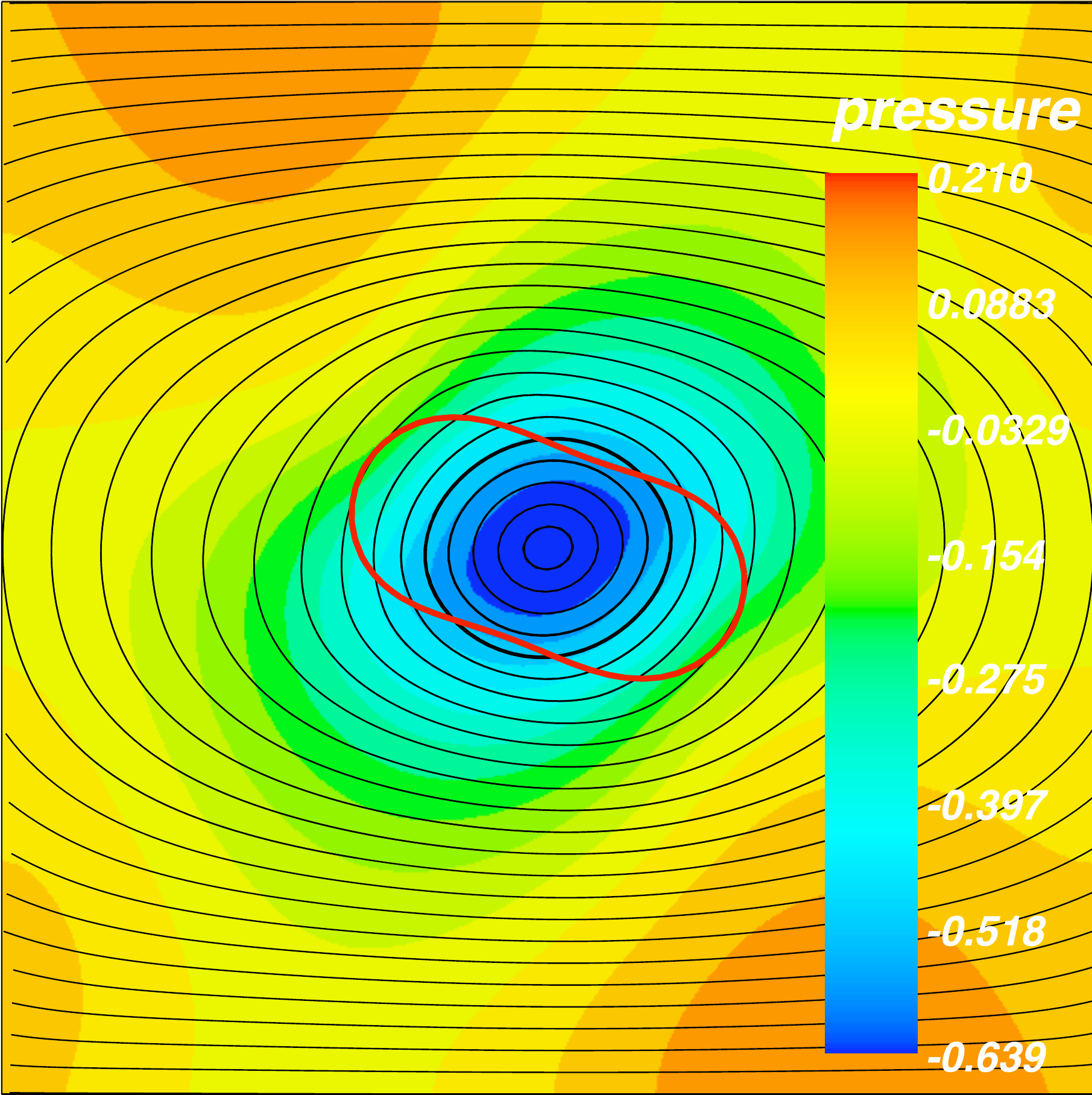}\includegraphics[width=3.5cm]{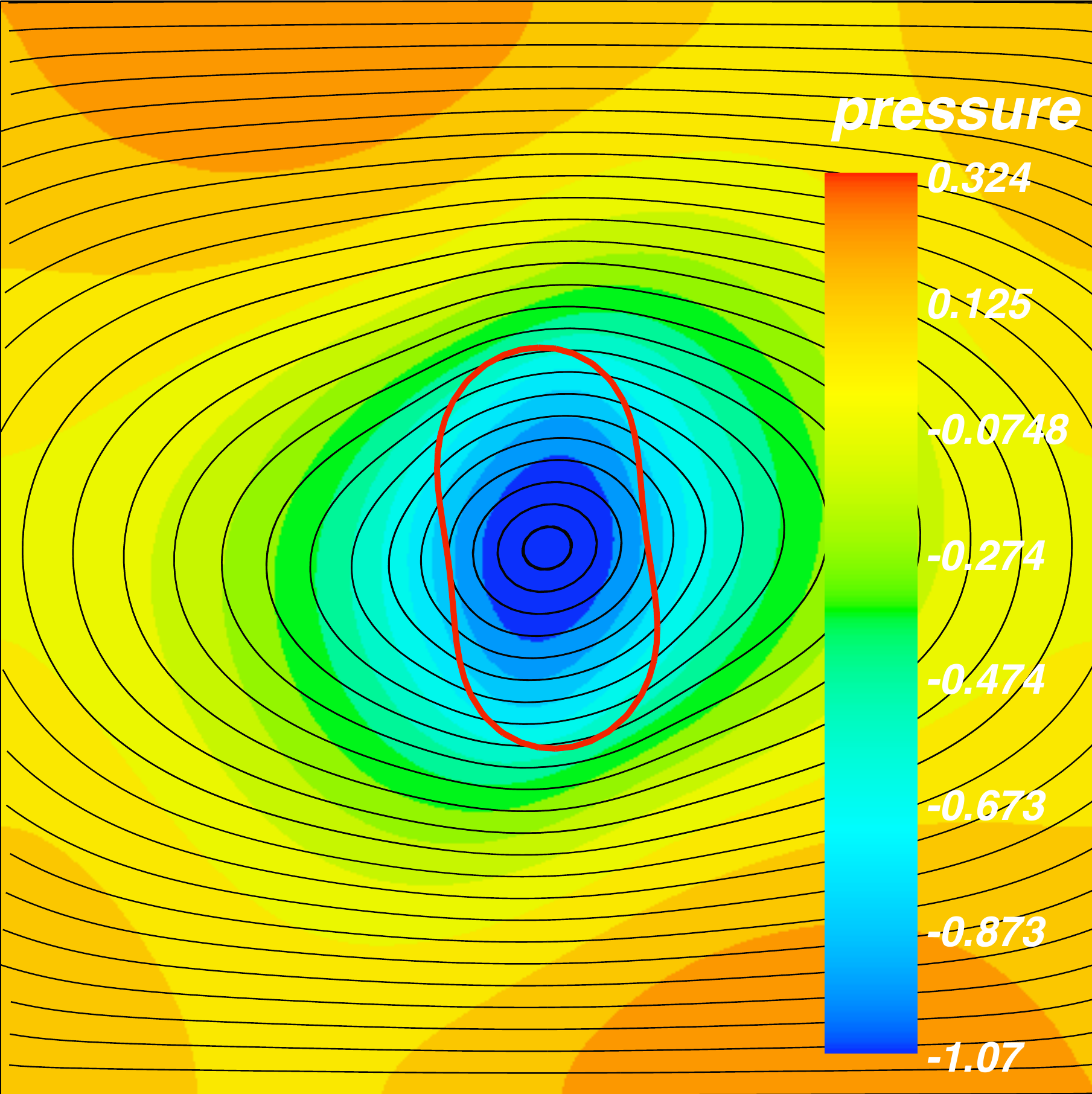}\includegraphics[width=3.5cm]{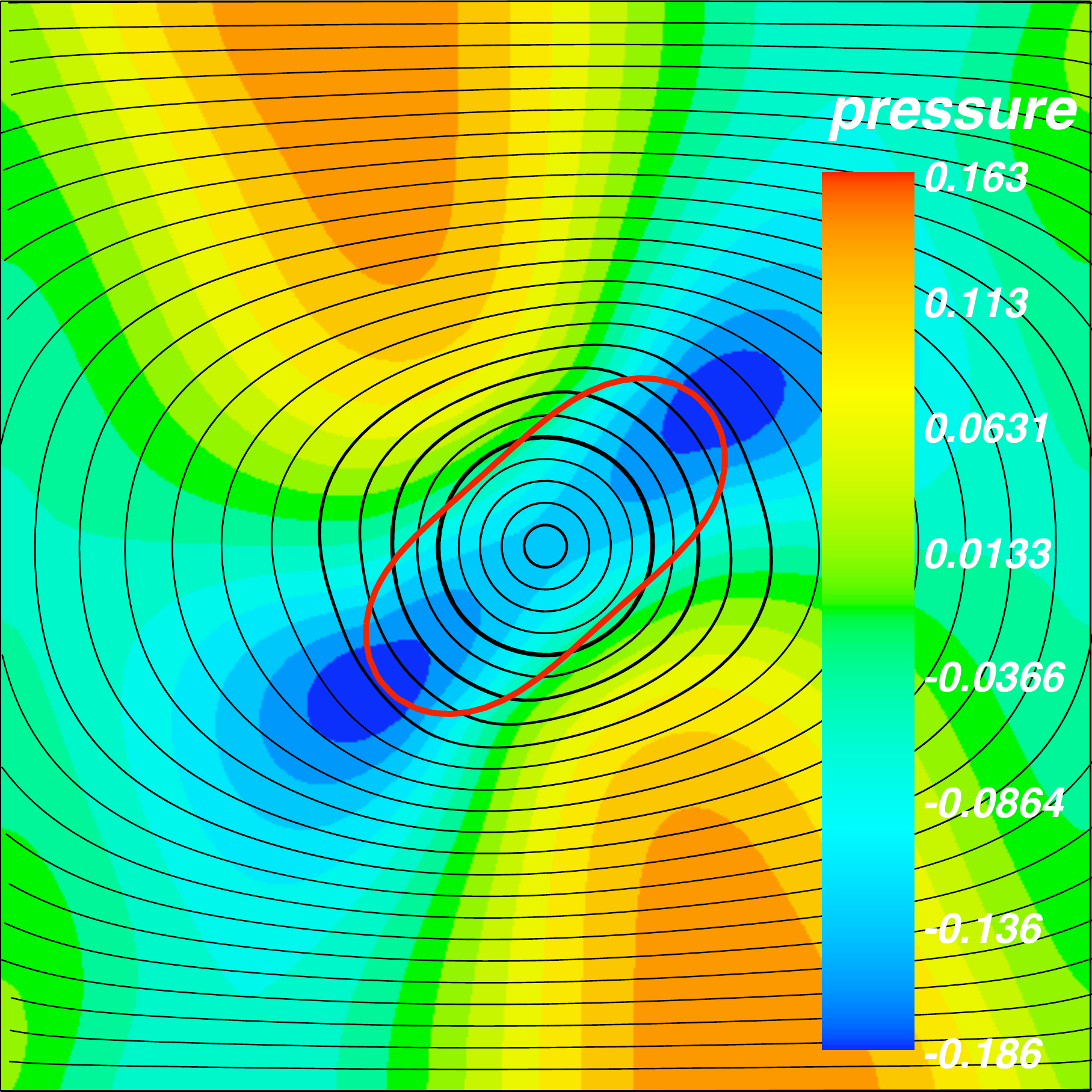}\includegraphics[width=3.5cm]{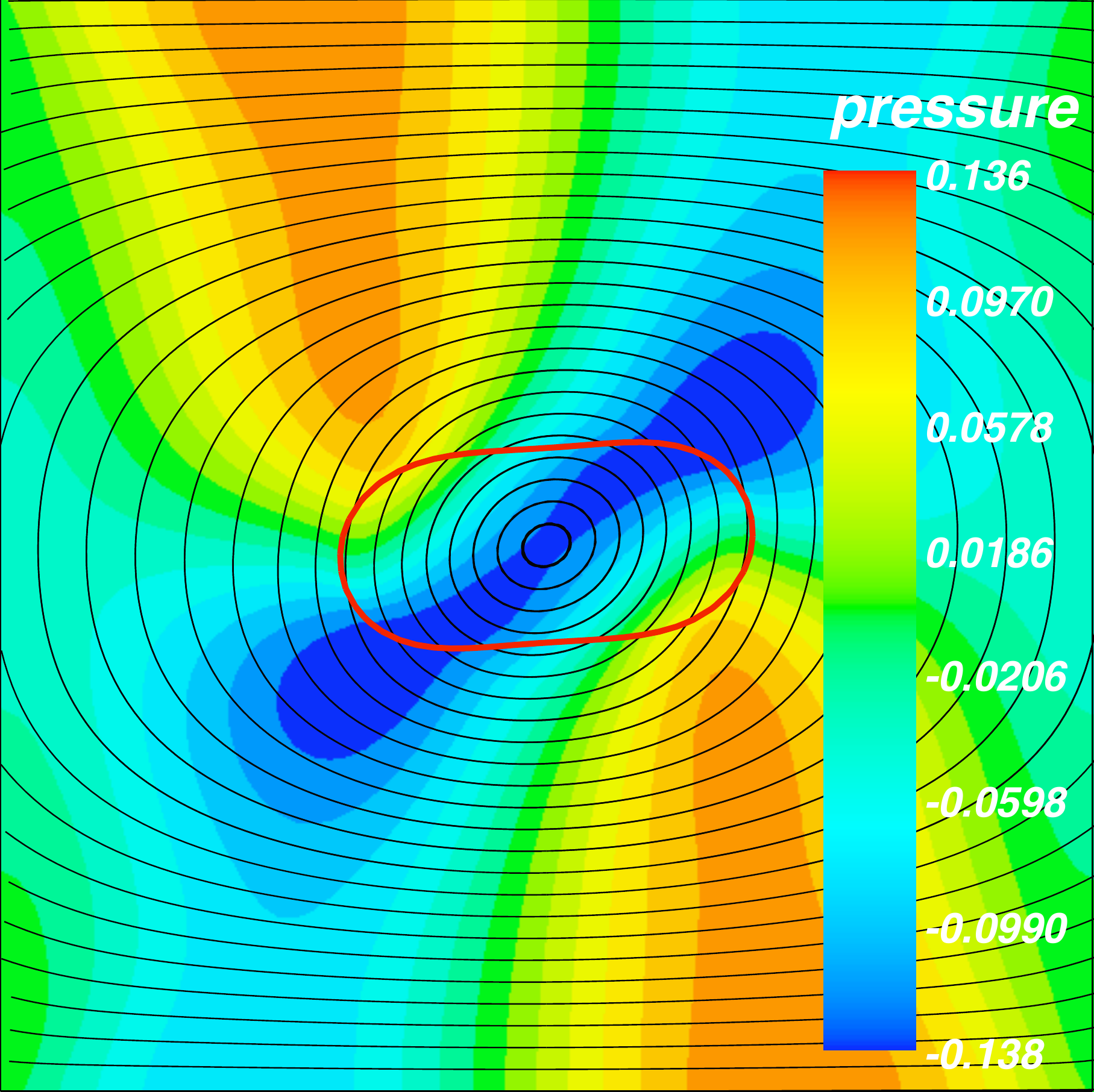}\includegraphics[width=3.5cm]{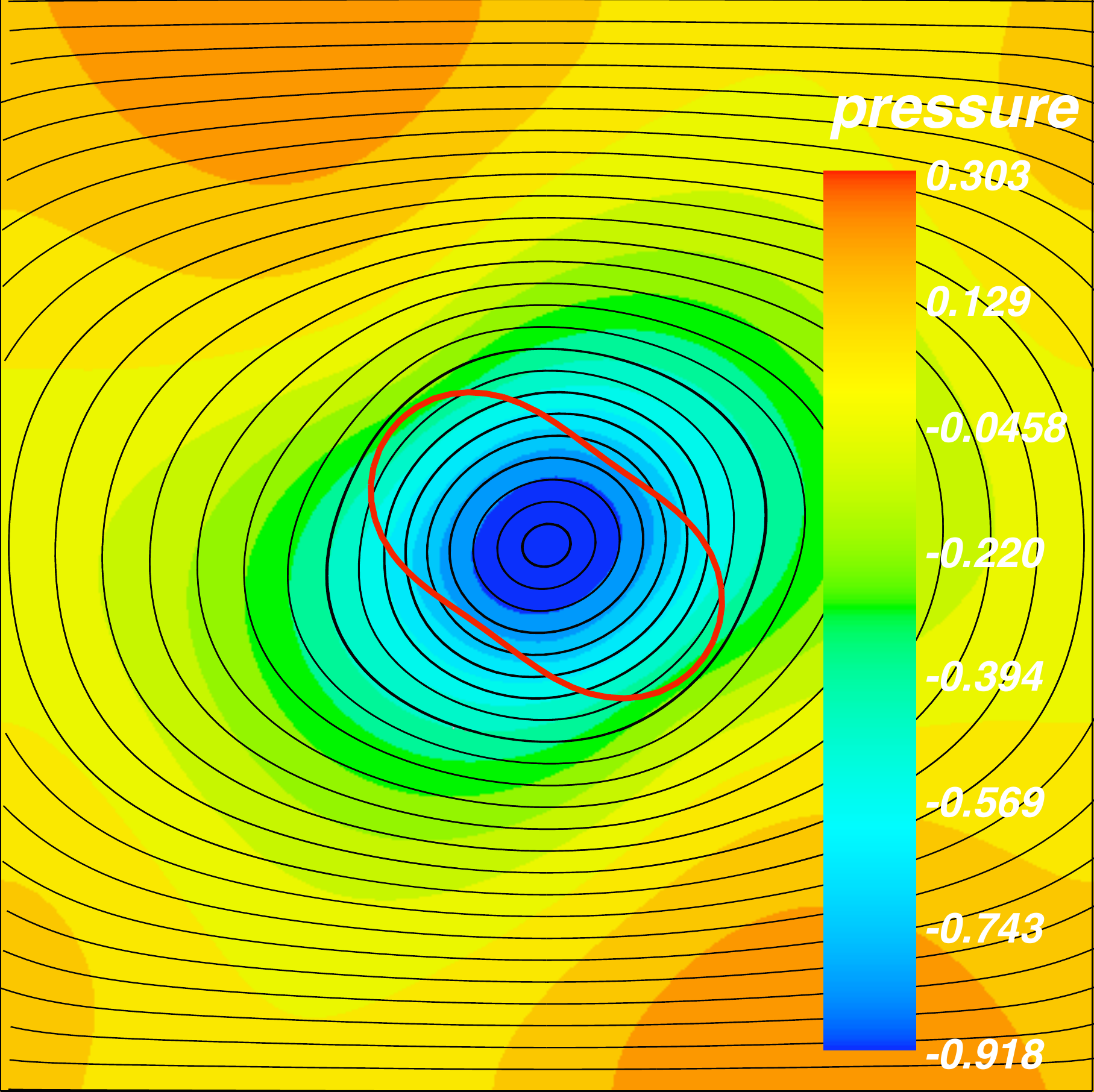}
\caption{\label{tumb2}Stream lines for the Tumbling motion of vesicle in a shear flow, at time $0.2$, $3$, $6$, $9$, $12$, $15$, $18$, $21$, $24$ and $27$. Viscosity contrast: $8$. $R_e=8.10^{-4}$. Colors stand for iso-pressure lines.}
\end{center}
\end{figure}

Figure \ref{tumb}  corresponds to the same parameters as before, but with a viscosity ratio $\lambda=8$ (the inner viscosity $8$ times larger than in the former test), we observe a tumbling motion.
It is interesting to see what happens if the external viscosity is lowered by a factor $8$. The viscosity contrast is still the same, but in this case  the external viscosity is lowered (rather than increasing the internal one, as done above).  Tumbling motion still prevails, but here the vesicle is more deformed than in the previous case (the peanut shape of the vesicle is more evident (Fig.\ref{tumb2})). This result is understood by noting that decreasing $\eta_{out}$ is equivalent to the previous situation provided that one increase $C_k$ (as well as $W_e$, while $R_e$ is reduced by a factor of $8$; actually we have multiplied $R_e$ by $8$ in this simulation, since we do not want to vary  the relative effect of inertia, as we have in mind the pure Stokes limit) by the same amount (a factor of 4), due to the similarity properties of the two situations (actually  the two situations are equivalent). Since $C_k$ is a measure of bending modes, its increase allows for more flexibility of the vesicle.

\begin{figure}
\centering
\mbox{\subfigure[Tank treading  for $\lambda=2$]{\includegraphics[width=14cm]{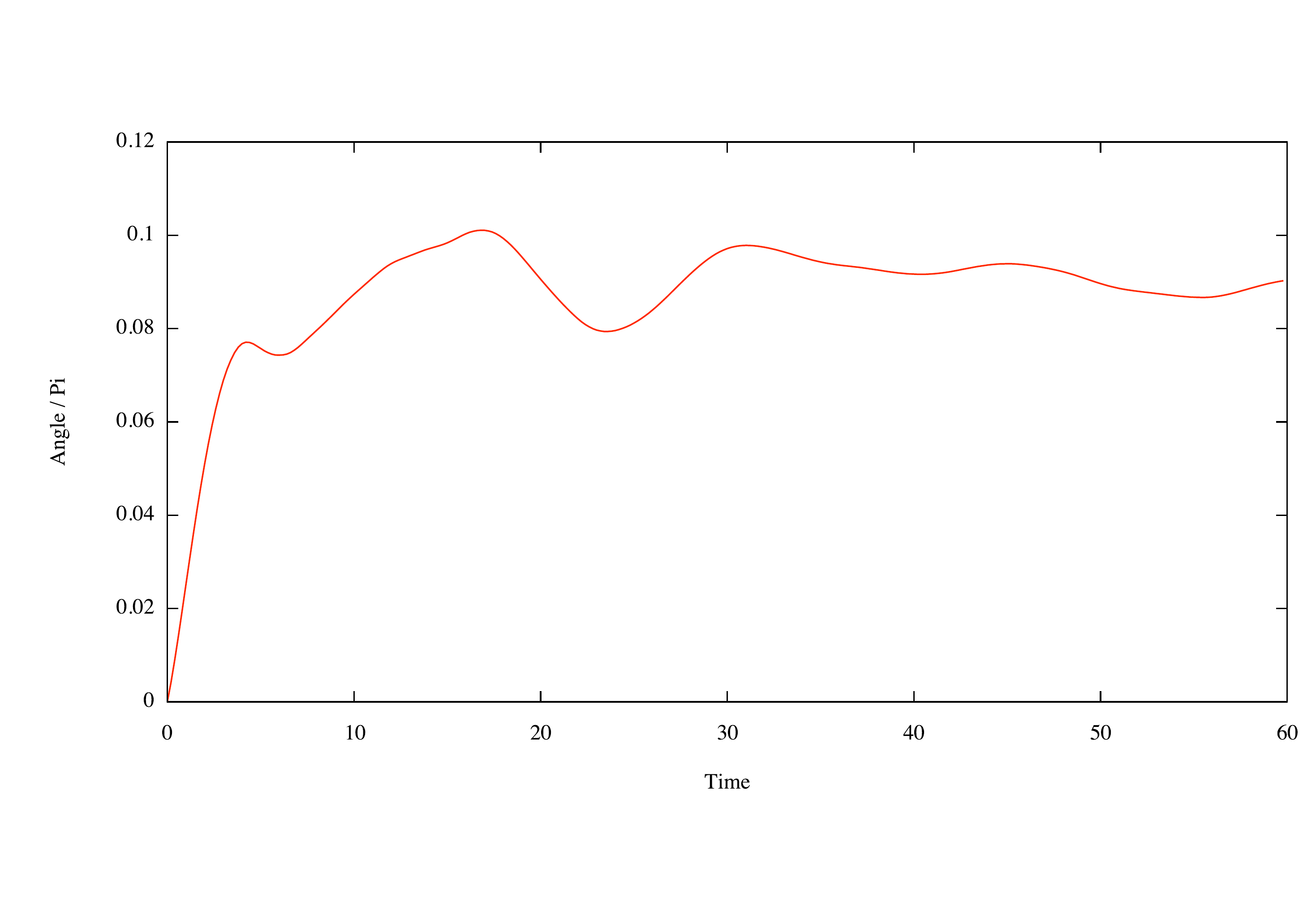}}}\\
\mbox{\subfigure[Vascillating breathing  for $\lambda=5.6$]{\includegraphics[width=7cm]{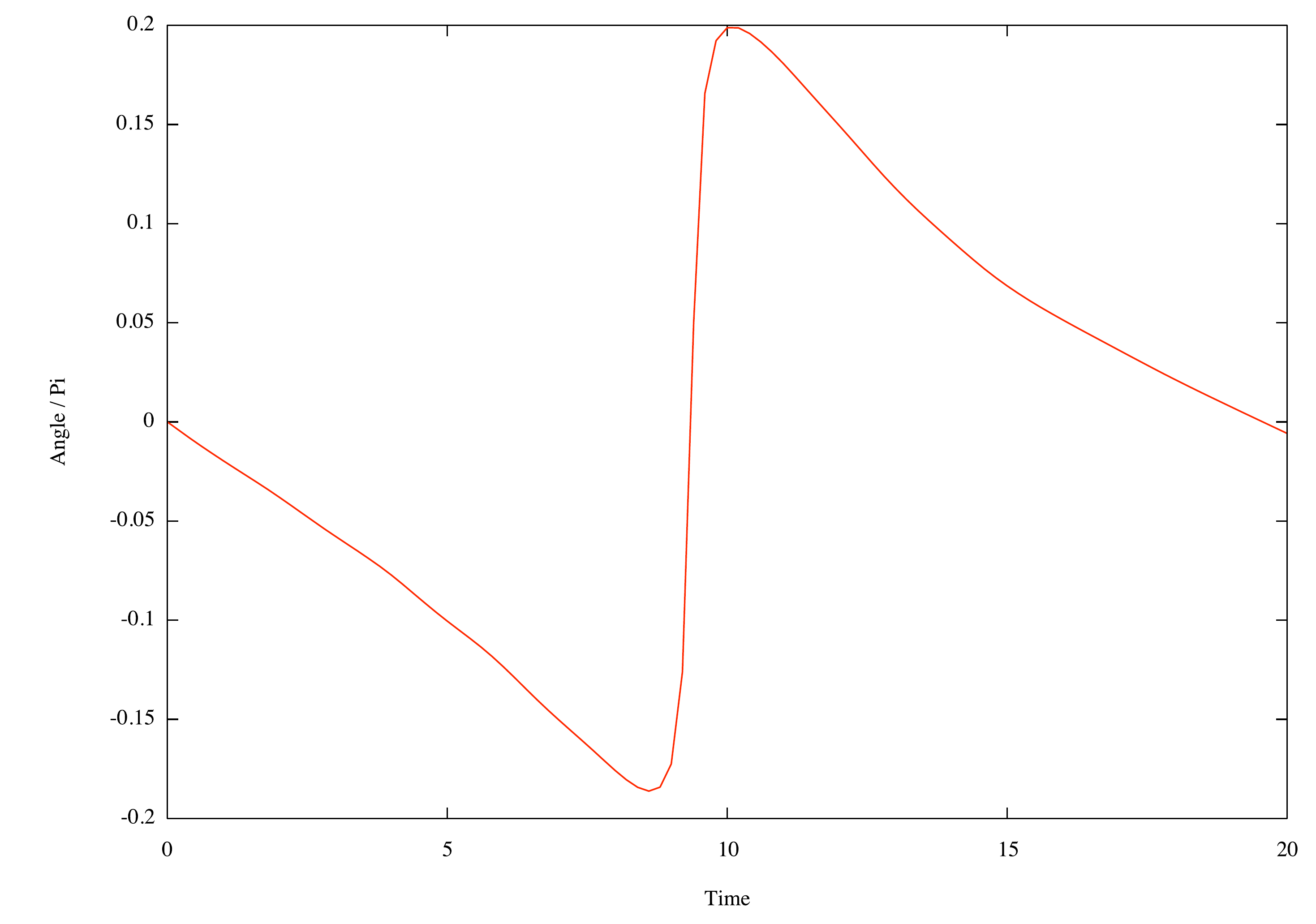}}\quad\subfigure[Tumbling for $\lambda=7$.]{\includegraphics[width=7cm]{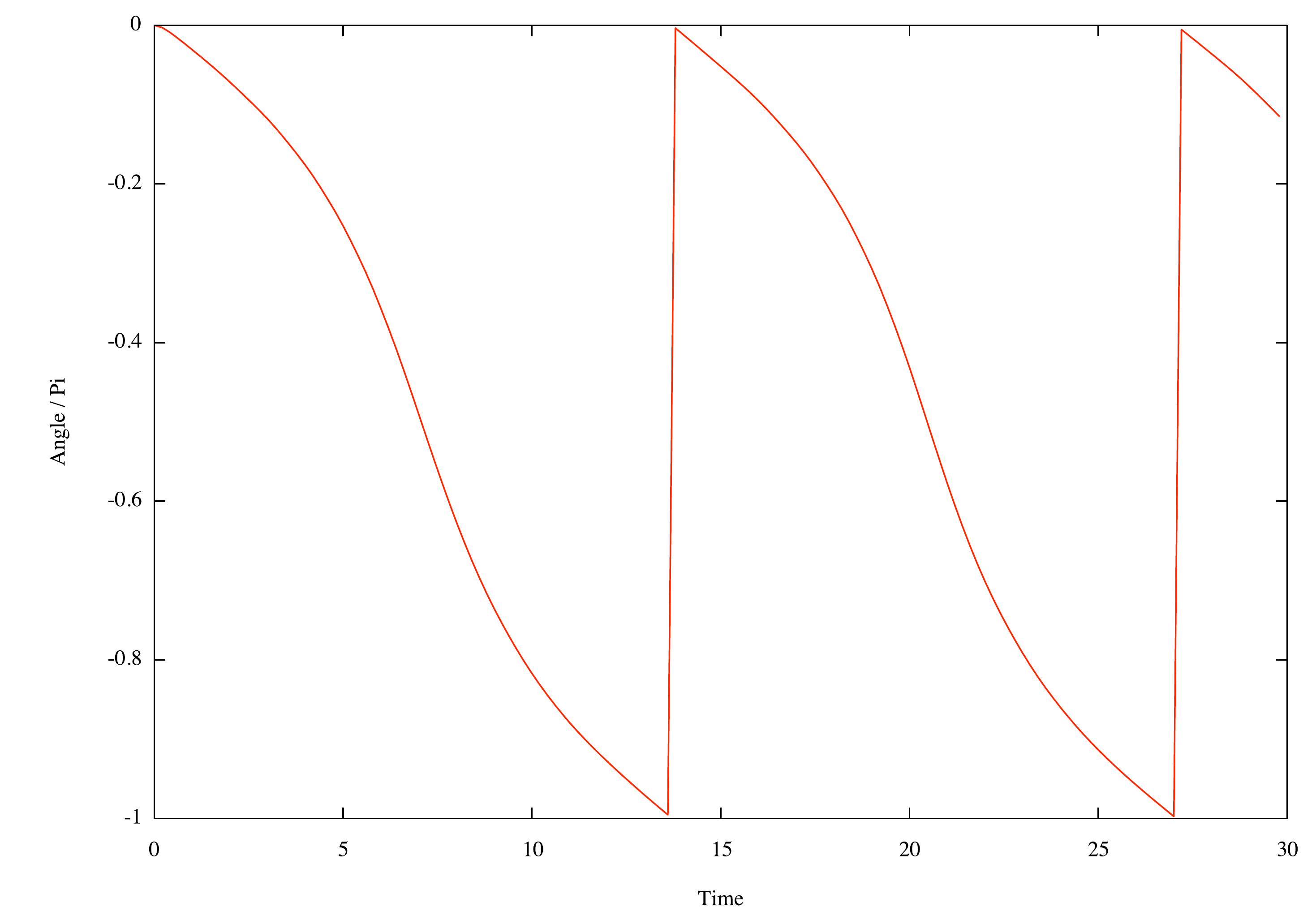}}}
\caption{\label{angle} Angle variation depending on the viscosity ratio, for $\nu=0.8$.}
\end{figure}

Figure \ref{angle} shows the variation of the angle of the vesicle main axis with respect to the flow direction (when the angle is zero this means that the vesicle is elongated and aligned along the flow direction). At low enough viscosity contrast the vesicle shows a tang-treading (Fig.\ref{angle}a)
At large enough $\lambda$ tumbling prevails (Fig.\ref{angle}c). We have found that at intermediate regime there is a vacillating-breathing regime as shown
on Fig.\ref{angle}b. Actually it can be shown from general arguments in {\it two dimensions} that for a fully inextensible vesicle, that a vacillating-breathing mode  can not take place in the small deformation regime (i.e. close to a circular shape where  only the second order mode is taken into account; the mode of a deformation about a circular shape is proportional to $e^{im\psi}$ where $\psi$ is the polar angle and $m$ is an integer. The second mode corresponds to $m=2$). On the one hand, at larger deformation the argument does not hold. In addition, recent numerical studies based on the boundary integral formulation \cite{Ghigliotti2009} did not observe the vacillating-breathing, but only as a transient. However, in that work the inextensibility condition was preserved up to a relative variation of about $10^{-3}$ for a resolution of $128^2$. If the variation is larger, then the vesicle may behave as a capsule, and in that case it has been recently shown that the vacillating-breathing mode precedes indeed the tumbling one on increasing $\lambda$. A systematic future study is needed in order to ascertain this more quantitatively.

\begin{figure}
\begin{center}
\includegraphics[width=10cm]{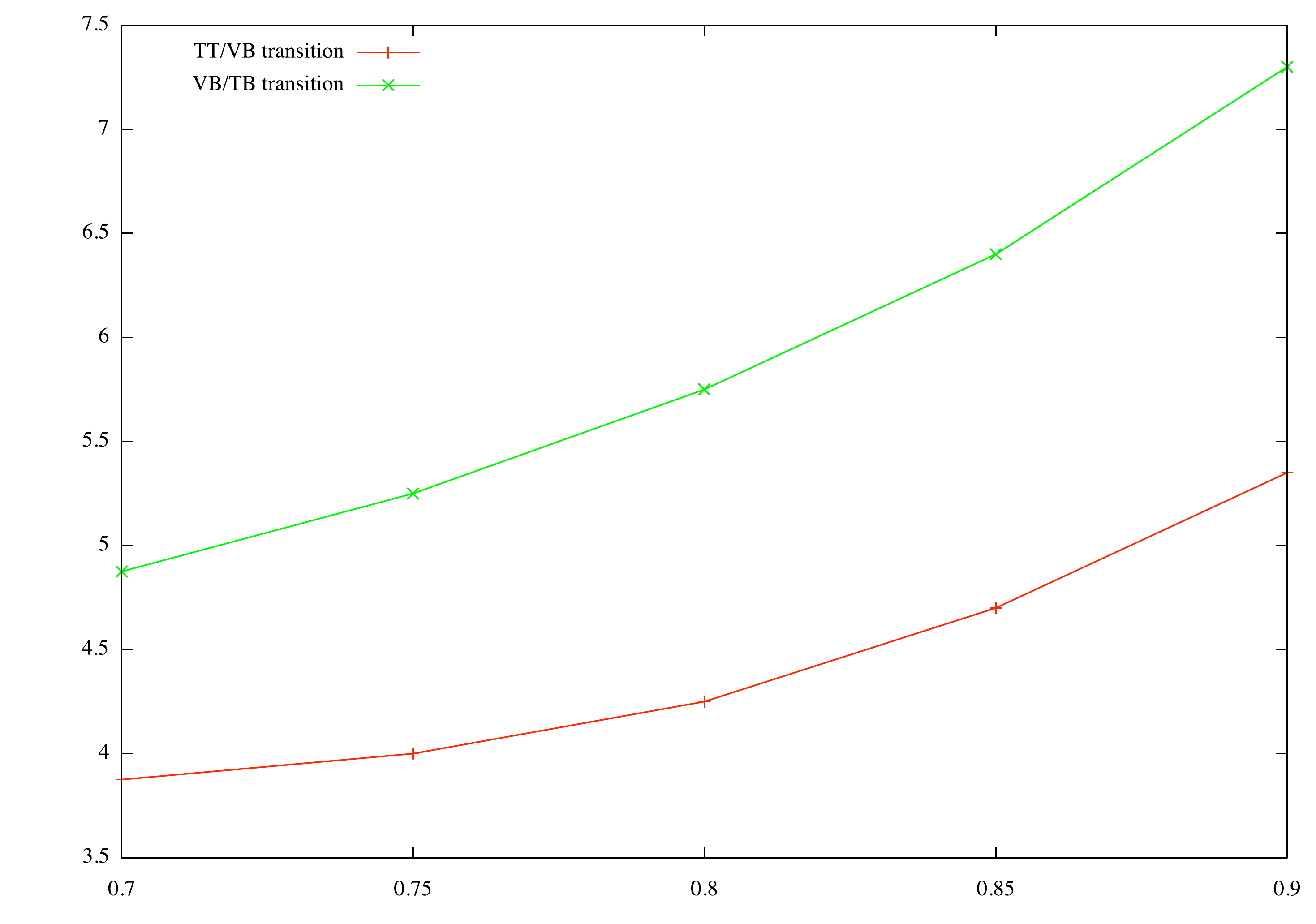}
\caption{\label{trans}Transition between different modes depending on the reduced volume  for a confinement of $0.3$ (confinement is defined as the ratio of the vesicle effective diameter (the diameter of a circle having the same perimeter as the actual vesicle) over the channel width.}
\end{center}
\end{figure}

Finally we draw the phase diagram representing the three types of motion in the plane of the reduced volume and the viscosity contrast, as has been done in Ref.\cite{Beaucourt2004}. The results are reported on Fig. \ref{trans}. The line of transition towards tumbling is quantitatively close to that reported in in Ref.\cite{Beaucourt2004}, albeit the present line is a bit lower. This can be attributed to confinement (in Ref. \cite{Beaucourt2004} the most quantitative results are obtained from the boundary integral formulation in an unbounded domain).


\section{Some bridges between the methods}
\label{bridge}
\subsection{Divergence + gradient form of the membrane forces}
\label{sec41}
Let us first show that the total membrane force (the bending  and elastic contribution) can be written as divergence of tensor plus a gradient
Let us start with the elastic  force (\ref{mforce}) and rewrite it as
$$\bF_m=\bnabla[E'(\ngvp)]\ngvp\ze-\div\left[E'(\ngvp)\frac\gvp\ngvp\right]\gvp\ze=:A-B$$
we use the following notations and elementary results from tensorial analysis:
\begin{itemize}
\item[(i)] For two vectors $a$ and $b$, $a\otimes b$ is the matrix with coefficient $a_ib_j$ on line $i$, column $j$.
\item[(ii)] This matrix is such that $(a\otimes b)c=(b\cdot c)a$; thus $n\otimes n$ is the projector on $n$.
\item[(iii)] For two vector fields $a$ and $b$, $\div(a\otimes b)=(\nabla a) b+(\div b)a$.
\item[(iv)] For a scalar function $f$ and a vector field $a$, $\nabla (fa)=a\otimes \nabla f+f\nabla a$.
\end{itemize}
The $B$ term in $F_m$ may be transformed by taking $a=\ze\gvp$ and $b=E'(\ngvp)\frac\gvp\ngvp$ in (iii) to get
$$B=\div\left(E'(\ngvp)\frac{\gvp\otimes\gvp}{\ngvp}\ze\right)-\nabla(\ze\gvp)E'(\ngvp)\frac\gvp\ngvp$$
Observe that
$$\nabla(\ze\gvp)=\gvp\otimes\nabla(\ze)+\ze\nabla^2\vp=\nabla(\ze)\otimes\gvp+\ze\nabla^2\vp$$
(in general $a\otimes b\neq b\otimes a$ but equality occurs when $a$ and $b$ are collinear). Moreover from (ii) one has
$$(\nabla(\ze)\otimes\gvp)\frac\gvp\ngvp=\ngvp\nabla(\ze)$$
and
$$\nabla^2\vp\frac\gvp\ngvp=\nabla\ngvp.$$
Thus $B$ may be written as
$$B=\div\left(E'(\ngvp)\frac{\gvp\otimes\gvp}{\ngvp}\ze\right)-E'(\ngvp)\nabla(\ngvp\ze)$$
and combined with $A$ so that $F_m$ reads
\begin{equation}
F_m=\nabla\left\{E'(\ngvp)\ngvp\ze\right\}-\div\left(E'(\ngvp)\frac{\gvp\otimes\gvp}{\ngvp}\ze\right)\label{mforcediv}
\end{equation}
This form of $F_m$ is interesting since the gradient term is useless when plugged as a source term to incompressible (Navier-)Stokes equation (it is absorbed by the pressure gradient). Another way consists in including
 the gradient term as the divergence of a diagonal tensor:
\begin{equation}
F_m=\div\left(E'(\ngvp)\ngvp(\I-\frac{\gvp\otimes\gvp}{\ngvp^2})\ze\right)\label{mforcediv3}
\end{equation}

The remaining divergence term may be grouped with the divergence term of the fluid equation, turning the fluid-structure problem into  the study of a fluid containing a viscoelastic band having the following constitutive law:
$$\sigma = -p\I+\eta(\nabla u+\nabla u^t)+E'(\ngvp)\frac{\gvp\otimes\gvp}{\ngvp}\ze$$
This form is interesting since in the particular case $E'(r)=\lambda r$ (which corresponds to an elastic interface with zero area surface at rest) we recover the stress tensor of a Korteweg fluid (setting $\Phi=Z(\vp/\eps)$ with $Z'=\zeta$). This remark has been used in \cite{CotMaiMil} to prove existence results for fluid-structure coupling problems formulated in Eulerian coordinates.

Of course this remark holds also for the phase-field formulation \cite{Biben2005}. In this case the membrane force is given by
\begin{equation}
F_m=\nabla(\xi\ngp)-\div\left(\xi\frac{\gp\otimes\gp}{\ngp}\right)=\div\left(\xi\ngp(\I-\frac{\gp\otimes\gp}{\ngp^2})\right)\label{mforcediv2}
\end{equation}
Note that $\I-\frac{\gp\otimes\gp}{\ngp^2}$ is the projector on the tangent subspace to the vesicle membrane.\\

\subsection{Link between $\xi$ and $\ngvp$}
\label{Link}

The first apparent difference between the level-set and the phase-field methods is that in one case (the level-set) we have a pure advection, while in the second one we enforce the values of $\phi$ to be $\pm 1$ inside and outside the vesicle by introducing a double well potential. Remember, however (see section \ref{Numerical}), that $\vp$ if it is initially a distance function, this properties is not preserved under dynamics. One has thus usually to resort to
a renormalization such as to keep it a distance function \cite{osherbook} (note that in section \ref{Numerical} we propose an alternative as compared to the usual method). Thus  this procedure is somewhat similar to forcing $\phi$ in the phase-field to remain constant in the bulk phases.
More precisely, first observe that introducing the function $Z$ such that $Z(r)=2\int_{-\infty}^r\zeta(s)ds-1$, the phase function $Z(\frac{\vp}\eps)$ has a behaviour similar to $\phi$: it is equal to $-1$ when $\vp<-\eps$, inside the vesicle, and to $1$ when $\vp>\eps$, outside the vesicle.

Disregarding the gradient terms in the two expressions (\ref{mforcediv}) and (\ref{mforcediv2}) of $F_m$, which play no role as stated above, we remark that thanks to the chain rule we have
$$\frac{\gvp\otimes\gvp}{\ngvp}\ze=\frac{\nabla Z(\frac{\vp}\eps)\otimes \nabla Z(\frac{\vp}\eps)}{|\nabla Z(\frac{\vp}\eps)|}$$
Thus in order to compare the formulations, regarding  the membrane forces, we must compare $\xi$ (see equation \ref{fconfig}) from the phase-field method with $E'(\ngvp)$ (see equation \ref{mforce}) of the level-set method. It is easy to prove  that if $\vp$ verifies a transport equation with divergence free velocity field, then $\ngvp$ verifies
$$\ngvp_t+u\cdot\nabla\ngvp=\ngvp t\cdot(t\cdot \nabla u)$$
and dividing by $\ngvp$ and multiplying by $T$ to obtain
$$(T\log\ngvp)_t+u\cdot\nabla(T\log\ngvp)=T t\cdot(t\cdot \nabla u)$$
i.e. $T\log\ngvp$ verifies the same equation as the postulated \cite{Biben2003,Biben2005} equation for $\xi$ (equation \ref{eqxi}). Of course we can not conclude that $\xi=T\log\ngvp$, since the velocity fields in the two methods should not coincide. Indeed the $\phi$ equation in the phase-field approach is not a simple transport equation as in level-set, and this fact affects the elastic forces and thus the dynamics. But up to these technical misfits, one may consider that $\xi$ is more or less proportional to the $\log$ of the local extension of the membrane instead of being simply proportional to this stretching. Another consideration is that, roughly speaking, and up to the Ginzburg-Landau intrinsic energy, the phase-field method can be considered as the level-set method where we choose as constitutive law
$$E(r)=T r(\log r-1)$$
in the elastic membrane energy.
\subsection{Divergence + gradient form of the curvature force}
The level-set curvature force (\ref{cforce}) (a similar form could be extracted in the phase-field formulation) can  be put into divergence form. After a lengthy  calculation (see Annex II) we can write $F_c^{LS}$ under divergence form $F_c^{LS}=\div \sigma^\eps_c$ with the tensor $\sigma^\eps_c$ given by
\begin{multline}\sigma^\eps_c=\left\{\ngvp\left[G(\kp)\I+\frac{\gvp}{\ngvp}\otimes\nabla G'(\kp)\right]\left(\I-\frac{\gvp\otimes\gvp}{\ngvp^2}\right)\right.\\\left.-G'(\kp)\left(\I-\frac{\gvp\otimes\gvp}{\ngvp^2}\right)D^2\phi\left(\I-\frac{\gvp\otimes\gvp}{\ngvp^2}\right)\right\}\ze\label{curvdiv}\end{multline}
In this expression the higher order term under the divergence contains an extension of the second fundamental form of the surface, to the whole space.

\subsection{Comparison of curvature forces with that obtained in Ref.\cite{Biben2005}}
The above derivation was general without specifying the form of the function $G$.
Let us take $G(r)=\frac\kappa2 r^2$ (this is the Helfrich form) in (\ref{cforce}) which now reads
\begin{equation*}
F_c^{LS}=\kappa\div\left[-\frac{\kp^2}2\frac\gvp\ngvp+\frac1\ngvp\prpo\left(\nabla[\ngp \kp]\right)\right]\ze\gvp
\end{equation*}
while the curvature force of (\ref{fconfig}) is
\begin{equation*}
F_c^{AF}=-\frac{\kappa}2\left\{\frac{c^3}{2}+t\cdot\nabla(t\cdot\nabla c)\right\}\gp
\end{equation*}
which is equivalent to that derived in Ref.\cite{Biben2005} (equation A13, for $c_0=0$, where $c_0$ is the spontaneous curvature considered in Ref.\cite{Biben2005}; recall that the definition of the curvature there has the opposite sign here).

In two dimension, $\prpo(v)=(t\cdot v)t$ thus the formula for $F_c^{LS}$ is identical to formula $F_c^{AF}$ of \cite{Biben2003,Biben2005}, up to a factor $2$ and the convention for the sign of curvature. Note that in Ref.\cite{Biben2005} it was not realized that the force can be written as a divergence, as shown here. Instead the force was in the numerical study substituted by an approximate expression, leading to  $F_c^{AF}$. To arrive to that expression in Refs. \cite{Biben2003,Biben2005} the property $\div t=0$ was used, which is not valid in general. Indeed, $t=\frac{\rot\phi}{\ngp}$ thus $\div t=-\rot\phi\cdot\frac{\nabla\ngp}{\ngp^2}=-t\cdot\frac{\nabla\ngp}{\ngp}\neq 0$. In ref. \cite{Biben2003,Biben2005} the curvature force  (equation (8) of \cite{Biben2005}), could be obtained from (6)  only to leading order of the interface width. The present study shows in fact that expression (8) is in fact exact. While this remark also holds in three dimension, the algebra to prove the coincidence of level-set and phase field expressions is somehow cumbersome and is therefore omitted here.

\subsection{Comparison of curvature forces with those of \cite{Jamet2008}}
In \cite{Jamet2008} the authors found the following elastic stress for the curvature stress (we use their notation):
$$\tau^C=-(\Phi-\div\mathcal{T})\otimes\nabla\vp-\mathcal{T}D^2\vp$$
where (considering the case $\mathcal{C}_0=0$, i.e. no spontaneous curvature) they set
$$\mathcal{T}=\alpha c(\vp)\left(\I-\frac{\gvp}{\ngvp}\otimes\frac{\gvp}{\ngvp}\right),$$
and $$\Phi=-\frac52c(\vp)^2\frac{\gvp}{\ngvp}-\frac{2\alpha C(\vp)}{\ngvp^2}\left[D^2\vp\gvp-\Delta\vp\gvp\right]$$
We are going to show that up to the cut-off $\zeta$ term which is included in the phase-field function, this is the same as the transpose of $\sigma_c^\eps$, in the case $G(r)=\alpha\frac{r^2}2$, which proves that the stresses are the same (recall that all geometric quantity express the same in terms of the phase-field or level-set equation). As a first step we write $\Phi$ in another way. Indeed we note that
after an  expansion of the curvature expression (which is $c=div(\gvp/\ngvp )$) we obtain
$$c(\phi)=\frac{1}{\ngvp}\left(\Delta\vp-\frac{D^2\vp\gvp\cdot\gvp}{\ngvp^2}\right).$$
One also  has
$$\Phi\cdot\frac\gvp\ngvp=-\frac{5\alpha}2c(\phi)^2+2\alpha c(\phi)^2=-\frac{\alpha}2c(\phi)^2$$
and thus
$$\Phi=(\Phi\cdot\frac\gvp\ngvp)\frac\gvp\ngvp+\left(\I-\frac{\gvp}{\ngvp}\otimes\frac{\gvp}{\ngvp}\right)\Phi=-\frac{\alpha}2c(\phi)^2\frac\gvp\ngvp-2\alpha\frac{c(\vp)}{\ngvp^2}\left(\I-\frac{\gvp}{\ngvp}\otimes\frac{\gvp}{\ngvp}\right)D^2\vp\gvp$$
The transpose of the level-set  stress is given by
\begin{multline}\left(\sigma^\eps_c\right)^T=\alpha\frac{c(\vp)^2}2\ngvp\left(\I-\frac{\gvp\otimes\gvp}{\ngvp^2}\right)+\alpha\left(\I-\frac{\gvp\otimes\gvp}{\ngvp^2}\right)\left(\nabla c(\vp)\otimes\gvp\right)\\-\alpha c(\vp)\left(\I-\frac{\gvp\otimes\gvp}{\ngvp^2}\right)D^2\phi\left(\I-\frac{\gvp\otimes\gvp}{\ngvp^2}\right)\label{stressLS}\end{multline}
Let us show that each term in the expression of  $\tau^C$ is hidden in this expression. Consider  the  term $\mathcal{T}D^2\vp$ which is included in the last term of (\ref{stressLS}), if we consider only the identity $I$ in the last projection term. The remainder is
\begin{equation}\alpha c(\vp)\left(\I-\frac{\gvp\otimes\gvp}{\ngvp^2}\right)D^2\phi\frac{\gvp\otimes\gvp}{\ngvp^2}\label{reste}\end{equation}
Now we focus on the term $\div\mathcal{T}\otimes\nabla\vp$. We first compute
\begin{multline*}
\div\mathcal{T}=\div\left[\alpha c(\vp)\left(\I-\frac{\gvp}{\ngvp}\otimes\frac{\gvp}{\ngvp}\right)\right]\\
=\alpha\left(\I-\frac{\gvp}{\ngvp}\otimes\frac{\gvp}{\ngvp}\right)\nabla c(\vp)+\alpha c(\vp)\div\left(\I-\frac{\gvp}{\ngvp}\otimes\frac{\gvp}{\ngvp}\right)\\
=\alpha\left(\I-\frac{\gvp}{\ngvp}\otimes\frac{\gvp}{\ngvp}\right)\nabla c(\vp)-\alpha c(\vp)\left(\I-\frac{\gvp}{\ngvp}\otimes\frac{\gvp}{\ngvp}\right)\frac{D^2\vp}\ngvp\frac\gvp\ngvp-\alpha c(\vp)^2\frac\gvp\ngvp\end{multline*}
Thus taking the tensorial product gives
\begin{multline*}
\div\mathcal{T}\otimes\gvp=\alpha\left(\I-\frac{\gvp}{\ngvp}\otimes\frac{\gvp}{\ngvp}\right)\nabla c(\vp)\otimes\gvp-\alpha c(\vp)\left(\I-\frac{\gvp}{\ngvp}\otimes\frac{\gvp}{\ngvp}\right)D^2\vp\left(\frac\gvp\ngvp\otimes\frac\gvp\ngvp\right)\\-\alpha c(\vp)^2\frac\gvp\ngvp\otimes\gvp\end{multline*}
The first term is exactly the second term of (\ref{stressLS}). The second term has the wrong sign to match (\ref{reste}). But combined with the new expression for $-\Phi\otimes\gvp$, namely
$$-\Phi\otimes\gvp=\frac{\alpha}2c(\phi)^2\frac\gvp\ngvp\otimes\gvp+2\alpha\frac{c(\vp)}{\ngvp^2}\left(\I-\frac{\gvp}{\ngvp}\otimes\frac{\gvp}{\ngvp}\right)D^2\vp\gvp\otimes\gvp$$
this gives the remaining terms, up to the extra term in the LS stress
$$ \alpha\frac{c(\vp)^2}2\ngvp\I$$
which is  a spherical tensor, thus not modifying the dynamics in an incompressible flow. Therefore the two curvature stress tensors are identical.

\section{Conclusion and forthcoming works}
\label{conclusion}
In this paper we investigated links between the phase-field and level-set modeling of immersed elastic membranes subject to curvature energy. We proved that these two formulations are equivalent from a theoretical point of view and validated the level-set formulation by recovering known behavior of phospholipid vesicles under shear flow. Undergoing works concern the generalization of our model to full membrane energy for the elastic membrane. Indeed while phospholipid vesicles only react to local change of area and curvature, red blood cells also react to shear in the tangent plane to their membrane. This is due to the spectrin network underneath. Therefore there is a need for the modeling of the full membrane energy of an immersed interface. This question is currently under investigation.

\section{Annexes}
\subsection{Annex I : differential of curvature energy}
Differentiating this energy with respect to $\vp$ gives
\begin{multline}
d\mathcal{E}_c(\vp)(\delta)=\int_\Om G'(\kp)\div\left(\frac\gd\ngvp-\frac{\gvp\cdot\gd}{\ngvp^3}\right)\ngvp\ze dx\\
+\int_\Om G(\kp)\frac{\gvp\cdot\gd}\ngvp\ze+G(\kp)\ngvp\zep\delta dx\label{eqcurv}
\end{multline}
Integration by part of the second term yields:
$$-\int_\Om G(\kp)\kp\ze\delta+G(\kp)\frac\gvp\ngvp\zep\gvp\delta+\nabla G(\kp)\cdot\frac\gvp\ngvp\ze\delta.$$
The second term cancels with the last one of equation (\ref{eqcurv}).
Denoting by $\prpo$ the linear projection operator on $\gvp^\perp$,
$$\prpo(\bv)=\bv-(\bv\cdot\frac{\gvp}{\ngvp})\frac{\gvp}{\ngvp}.$$
we thus have
$$d\mathcal{E}_c(\vp)(\delta)=\!\!\int_\Om G'(\kp)\div\left(\frac{\prpo(\gd)}\ngvp\right)\ngvp\ze-G(\kp)\kp\ze\delta-\nabla G(\kp)\cdot\frac\gvp\ngvp\ze\delta dx\!\!\!$$
which thanks to the expression for $\kp$ also reads
$$d\mathcal{E}_c(\vp)(\delta)=\int_\Om G'(\kp)\div\left(\frac{\prpo(\gd)}\ngvp\right)\ngvp\ze-\div(G(\kp)\frac\gvp\ngvp)\ze\delta dx$$
Since $\prpo(\gd)\cdot\gvp=0$ the first term integrates by part into
$$-\int_\Om\nabla\left[\ngvp G'(\kp)\right]\cdot\prpo(\gd)\frac1\ngvp\ze=-\int_\Om\prpo\left[\ngvp\nabla G'(\kp)\right]\cdot\frac{\gd}{\ngvp}\ze dx$$
where we used the symmetry property of the projection on $\gvp^\perp$. Integrating by parts once again, there holds
$$d\mathcal{E}_c(\vp)(\delta)=\int_\Om\div\left[-G(\kp)\frac\gvp\ngvp+\frac1\ngvp\prpo\left(\nabla[\ngvp G'(\kp)]\right)\right]\ze\delta dx$$

\subsection{Annex II : divergence form of curvature energy}
To get that divergence form we start from (\ref{cforce}) that we recall for convenience
\begin{equation*}
F_c=\div\left[-G(\kp)\frac\gvp\ngvp+\frac1\ngvp\prpo\left(\nabla[\ngvp G'(\kp)]\right)\right]\ze\gvp
\end{equation*}
and first use the tensor identity (iii) in \ref{sec41} which gives
\begin{multline*}
F_c=\div\left[\left\{-G(\kp)\frac{\gvp\otimes\gvp}\ngvp+\frac1\ngvp\gvp\otimes\prpo\left(\nabla[\ngvp G'(\kp)]\right)\right\}\ze\right]\\-\nabla\left(\ze\gvp\right)\times\left(-G(\kp)\frac\gvp\ngvp+\frac1\ngvp\prpo\left(\nabla\ngvp G'(\kp)\right)\right)
\end{multline*}
that we denote by $\div A-B$. Note that $\times$ denotes a matrix-vector product in the above formula.
Working on the second term and using (iv) of \ref{sec41}, $\nabla\left(\ze\gvp\right)=\gvp\otimes\nabla\ze+\ze D^2\vp=\zep\gvp\otimes\gvp+\ze D^2\vp=\nabla\ze\otimes\gvp+\ze D^2\vp$. Moreover, recall that
$$\frac{D^2\vp\gvp}{\ngvp}=\nabla\ngvp\text{ and }\left(\nabla\left(\ze\right)\otimes\gvp\right)\prpo(u)=0$$
for any $u$ using (ii) from \ref{sec41}. Thus
\begin{multline*}
 B=-G(\kp)\ngvp\nabla\ze-G(\kp)\ze\nabla\ngvp+\frac{D^2\vp}{\ngvp}\prpo\left(\nabla\ngvp G'(\kp)\right)\ze\\
=-G(\kp)\nabla\left(\ngvp\ze\right)+\frac{D^2\vp}{\ngvp}\prpo\left(\nabla\ngvp G'(\kp)\right)\ze
\end{multline*}
From the definition of projector, $$\prpo(u)=\left(\I-\frac{\gvp\otimes\gvp}{\ngvp^2}\right)u$$
and by using $u\otimes Av=(u\otimes v)A^T$,
$$\nabla\left(\frac\gvp\ngvp\right)=\left(\I-\frac{\gvp\otimes\gvp}{\ngvp^2}\right)\frac{D^2\vp}{\ngvp}\qquad \nabla\left(\frac\gvp\ngvp\right)^T=\frac{D^2\vp}{\ngvp}\left(\I-\frac{\gvp\otimes\gvp}{\ngvp^2}\right)$$
Let us try to write $B$ as a divergence term minus a remainder term. To start with,
$$B=-\nabla\left(G(\kp)\ngvp\ze\right)+G'(\kp)\nabla\kp\ngvp\ze+\nabla\left(\frac\gvp\ngvp\right)^T\nabla\left(\ngvp G'(\kp)\right)\ze$$
and computing $\nabla\kp$ leads to
$$\nabla\kp=\nabla\left(\div\frac\gvp\ngvp\right)=\div\left[\nabla\left(\frac\gvp\ngvp\right)^T\right]$$
Therefore
\begin{align*}
 B&=-\nabla\left(...\right)+\left\{G'(\kp)\ngvp\div\nabla\left(\frac\gvp\ngvp\right)^T+\nabla\left(\frac\gvp\ngvp\right)^T\nabla\left(\ngvp G'(\kp)\right)\right\}\ze\\
&=-\nabla\left(...\right)+\div\left(G'(\kp)\ngvp\nabla\left(\frac\gvp\ngvp\right)^T\right)\ze\\
&=-\nabla\left(...\right)+\div\left(G'(\kp)\ngvp\nabla\left(\frac\gvp\ngvp\right)^T\ze\right)-G'(\kp)\ngvp\nabla\left(\frac\gvp\ngvp\right)^T\nabla\left(\ze\right)
\end{align*}
but the last term is zero thanks to the expression of $\nabla\left(\frac\gvp\ngvp\right)^T$ with the projector on $\gvp^\perp$ and the fact that $\nabla\left(\ze\right)$ is colinear to $\gvp$. Writing the gradient term as the divergence of a diagonal tensor we get a first expression in divergence form:
\begin{multline*}F_c=\div\left[\left\{-G(\kp)\ngvp\left(\I-\frac{\gvp\otimes\gvp}{\ngvp^2}\right)+\frac\gvp\ngvp\otimes\prpo\nabla\left(\ngvp G'(\kp)\right)\right.\right.\\\left.\left.-G'(\kp)\ngvp\nabla\left(\frac\gvp\ngvp\right)^T\right\}\ze\right]
\end{multline*}
We may arrange terms further by using $u\otimes Av=(u\otimes v)A^T$,
$$\frac{\gvp}\ngvp\otimes\prpo(v)=\frac{\gvp}\ngvp\otimes\left(\left(\I-\frac{\gvp\otimes\gvp}{\ngvp^2}\right)v\right)=\left(\frac{\gvp}\ngvp\otimes v\right)\left(\I-\frac{\gvp\otimes\gvp}{\ngvp^2}\right)$$
which with $v=\nabla\ngvp G'(\kp)$ gives
\begin{align*}
 \frac\gvp\ngvp&\otimes\prpo\left(\nabla\ngvp G'(\kp)\right)=\left(\frac{\gvp}\ngvp\otimes \nabla\left(\ngvp G'(\kp)\right)\right)\left(\I-\frac{\gvp\otimes\gvp}{\ngvp^2}\right)\\
&=G'(\kp)\left(\frac{\gvp}\ngvp\otimes\frac{D^2\vp}{\ngvp}\right)\left(\I-\frac{\gvp\otimes\gvp}{\ngvp^2}\right)+\gvp\otimes\nabla G'(\kp)\left(\I-\frac{\gvp\otimes\gvp}{\ngvp^2}\right)
\end{align*}
and $$-G'(\kp)\ngvp\nabla\left(\frac\gvp\ngvp\right)^T=-G'(\kp) D^\vp\left(\I-\frac{\gvp\otimes\gvp}{\ngvp^2}\right)$$
Finally we get the expression $F_c=\div\sigma^\eps_c$ with
\begin{multline}\sigma^\eps_c=\left\{G(\kp)\ngvp\left(\I-\frac{\gvp\otimes\gvp}{\ngvp^2}\right)+\left(\gvp\otimes\nabla G'(\kp)\right)\left(\I-\frac{\gvp\otimes\gvp}{\ngvp^2}\right)\right.\\\left.-G'(\kp)\left(\I-\frac{\gvp\otimes\gvp}{\ngvp^2}\right)D^2\phi\left(\I-\frac{\gvp\otimes\gvp}{\ngvp^2}\right)\right\}\ze\end{multline}
that we can arrange like in (\ref{curvdiv}).

\section*{Acknowledgements}
This work was supported by the French Ministry of Education through ANR MOSICOB and by University Joseph Fourier by PPF DYSCO. E.M. was also supported by ANR COMMA. C.M. Acknowledges financial support from CNES.

\bibliography{biblio_bis}
\end{document}